\documentstyle[12pt,aaspp,psfig]{article}
\slugcomment{\shortstack[r]{resubmitted 1st November 2002}}
\begin{document}

\newcommand\Msun {M_{\odot}\ }

\title{VLT/UVES Abundances in Four Nearby Dwarf Spheroidal 
Galaxies\footnotemark[1]
II: Implications for Understanding Galaxy Evolution}
\vskip0.5cm

\author{{\bf Eline Tolstoy,}}
\affil{Kapteyn Institute, University of Groningen, PO Box 800, 9700AV
Groningen, the Netherlands}

\author{{\bf Kim A. Venn,}}
\affil{Macalester College, Saint Paul, MN 55105, USA\\
University of Minnesota, 116 Church Street S.E., Minneapolis, MN 55455, USA}

\author{{\bf Matthew Shetrone,}}
\affil{University of Texas, McDonald Observatory, HC75 Box 1337-L,
Fort Davis, Tx 79734 USA}

\author{{\bf Francesca Primas,}}
\affil{European Southern Observatory, Karl-Schwarzschild str. 2,
D-85748 Garching bei M\"{u}nchen, Germany}

\author{{\bf Vanessa Hill,}}
\affil{Observatoire de Paris-Meudon, GEPI, 2 pl.  Jules Janssen, 92195
Meudon Cedex, France}

\author{{\bf Andreas Kaufer \& Thomas Szeifert}}
\affil{European Southern Observatory, Alonso de Cordova 3107,
Santiago, Chile}

\footnotetext[1]{Based on Ultraviolet-Visual Echelle Spectrograph
observations collected at the European Southern Observatory, proposal
numbers 66.B-0320, 65.N-0378, 62.N-0653 and 61.A-0275}

\newpage

\begin{abstract}

We have used the Ultra-Violet Echelle Spectrograph (UVES) on Kueyen
(UT2) of the VLT to take spectra of 15 individual red giant stars in
the centers of four nearby dwarf spheroidal galaxies: Sculptor,
Fornax, Carina and Leo~I. We measure the abundance variations of
numerous elements in these low mass stars with a range of ages
(1$-$15~Gyr old). This means that we can effectively measure the
chemical evolution of these galaxies {\it with time}.

Our results show a significant spread in metallicity with age, but an
overall trend consistent with what might be expected from a closed (or
perhaps leaky) box chemical evolution scenario over the last
10$-$15Gyr. We make comparisons between the properties of stars
observed in dSph and in our Galaxy's Disk and Halo, as well as
Globular Cluster populations in our Galaxy and in the Large Magellanic
Cloud. We also look for the signature of the earliest star formation
in the Universe, which may have occurred in these small systems.

We notice that each of these galaxies show broadly similar abundance
patterns for all elements measured. This suggests a fairly uniform
progression of chemical evolution with time, despite quite a large
range of star formation histories. It seems likely that these galaxies
had similar initial conditions, and evolve in a similar manner with
star formation occurring at a uniformly low rate, even if at different
times.  With our accurate measurements we find
evidence for small variations in abundances which seem to be
correlated to variations in star formation histories between different
galaxies. The $\alpha$-element abundances suggest that dSph chemical
evolution has not been affected by very high mass stars
($>15-20\Msun$).

The abundance patterns we measure for stars in dwarf spheroidal
galaxies are significantly different from those typically observed in
the disk, bulge and inner-halo of our Galaxy.  This means that, as far
as we can tell from the (limited) data available to date it is
impossible to construct a significant fraction of our disk, inner-halo
or bulge from {\it stars} formed in dwarf spheroidal galaxies such as
we see today which subsequently merged into our own.  Any merger
scenario involving dSph has to occur in the very early Universe whilst
they are still gas rich, so the majority of mass transfer is gas, and
few stars.

\end{abstract}

\keywords{GALAXIES: ABUNDANCES GALAXIES: DWARF GALAXIES: INDIVIDUAL 
(SCULPTOR, FORNAX, CARINA, LEO I) STARS: ABUNDANCES}

\newpage

\section{Introduction}

With High Resolution Spectrographs on 8m class telescopes, such as
UVES on VLT/UT2, we can take detailed spectra of individual stars in
nearby dwarf galaxies and start to seek answers to detailed questions
about the enrichment {\it history} of a variety of different elements
within galaxies other than our own.  It is critical to our
understanding of galaxy evolution to accurately measure how the
fraction of heavy elements in a galaxy builds up {\it over time}. It
is impossible to accurately analyze a Color-Magnitude Diagram (CMD)
and extract a unique star formation history (SFH) without some
independent information on the metallicity of the stars and how this
may have changed with time.  The oldest stars we find in the nearby
Universe are of an age similar to, if not greater than, the look-back
time of the most distant galaxies found in the highest redshift
galaxies known.

There are many theories as to how the chemical enrichment of
subsequent generations of stars may occur in dwarf galaxies ({\it
e.g.}, Silk, Wyse \& Shields 1987; Pagel 1997), and there are also a
significant number of measurements of nebular abundances and young
stars today which undoubtedly carry the imprint of past enrichment
patterns ({\it e.g.}, Matteucci \& Tosi 1985; Pagel \&
Tautvai\v{s}ien\.e 1998; Venn {\it et al.} 2001).  However, extracting
unique chemical evolution scenarios from these data is very difficult.
The only way to accurately determine absolute abundances at different
ages in the history of a galaxy is to take high resolution spectra of
old low mass stars, with a range in age (color and luminosity).  With
extra-galactic stars in nearby galaxies we are forced to look at the
brightest stars we can, which are Red Giant Branch (RGB) stars which
were formed during the last 1$-$15~Gyr.  The abundances of different
elements provide an indication of the number and type of Supernovae
enrichments which have proceeded the formation of these stars. Thus we
can measure the evolution of these galaxies in great detail from the
earliest times up to the epoch of last star formation.  From previous
studies there are well established techniques for detailed abundance
analysis ({\it e.g.}, Shetrone, Bolte \& Stetson 1998; Shetrone,
C\^{o}t\'{e} \& Sargent 2001 [SCS01]; Hill {\it et al.} 2000 [H00])
which we can apply to the same type of stars in our sample of southern
dwarf spheroidals.

Dwarf spheroidal galaxies (dSph) are intriguing systems. They are our
nearest neighbors and so we can study them in great detail.  Within
the Local Group they are mostly satellites of larger galaxies such as
our Galaxy and M~31 (e.g., van den Berg 2000).  
Originally all dSph were believed to be
essentially old systems similar to globular clusters, although there
were hints to the contrary from the RR~Lyrae stars and anomalous
Cepheids which suggested that their underlying stellar populations
were fundamentally different from those of globular clusters ({\it
e.g.}, Baade 1963; Zinn 1980).  Zinn suggested that the difference was
due to the dSph systems being on {\it average} younger than globular
clusters.  Subsequent deep photometric observations of the
main-sequence turnoffs (MSTOs) in these galaxies confirmed the
presence of intermediate age populations in some galaxies, and always
an age spread at the earliest times, completely unlike most globular
clusters. The dSph are typically small, faint and diffuse, non-nucleated
systems, and yet they have complex SFHs.  For most of these systems
the SFHs have been accurately measured from main sequence turnoff
photometry.  These galaxies have typically no associated ISM, no HI
gas (in the center), although there have been tentative detections of
gas in the out lying regions of a few nearby dSph ({\it e.g.},
Carignan {\it et al.} 1998).

Despite considerable efforts with 4m class telescopes, many of the
most basic questions remain unanswered for most of the Local Group
dSph: How and when were they formed? Are they left over building
blocks of our Galaxy? How has environment effected their evolution?
To have a hope of addressing these complex and difficult questions we
need to first answer such mundane ones as: What is the total range in
[Fe/H] for stars within a given dSph?  Does the [Fe/H] match that of
our halo populations (Globular clusters or field stars)?  Do the
elemental ratios (particularly [$\alpha$/Fe] and [r- and s-
process/Fe] ratios) in dSph stars track those seen in the metal
poor field or the globular clusters? In addition, there is strong
evidence for a difference in the ratios of C, N, O, Na, Mg and Al to
Fe between field population~II giants and globular cluster giants
({\it e.g.}, Pilachowski {\it et al.} 1996; Shetrone 1996). This
difference is presumably related to the extent to which deep dredging
occurs in giants, but if (and how) this is ultimately caused by
variations in environment between globular cluster and field stars
remains a mystery.  However, if the dSph are ``globular-cluster-like''
in this regard, it would present a problem for the idea that the
Galactic halo is composed of dispersed dSph galaxies ({\it e.g.}, Zinn
1993; Shetrone 1998).

Detailed studies of the abundance patterns of stars in dSph can
constrain the effects of discrete star formation episodes and
different SFH on chemical evolution (e.g, McWilliam 1997).  UT2+UVES
provides a unique opportunity (in the southern hemisphere) to extend
detailed stellar abundance work to stars in external galaxies.  This
kind of study has only been possible so far with HIRES on Keck ({\it e.g.},
Shetrone, Bolte \& Stetson 1998; SCS01), where it has been clearly
shown that detailed abundance analysis of stars in nearby galaxies is
feasible and can place important constraints on metallicity
dispersions and what they mean for the metallicity evolution of nearby
galaxies. We looked at samples of stars in four nearby dwarf
spheroidal galaxies (dSphs): Sculptor, Fornax, Carina and Leo~I.  We
also looked at four stars in three well studied stars in Galactic
globular clusters (M30, M55, M68), to check that at least our
calibrations put us on the same abundance scale as previous studies.
Globular clusters also serve as an interesting comparison to the
observations of stars in dSph.

In Sculptor and Fornax there are recently published low resolution
Ca~II triplet (CaT) spectra of a sample of RGB stars (Tolstoy et
al. 2001; hereafter T01).  These UVES data thus provide a detailed
check of the CaT method which has been used to estimate iron
abundances in RGB stars, based on calibrations in the Galaxy ({\it e.g.},
Armandroff \& Da Costa 1991; Rutledge, Hesser \& Stetson 1997).

From our high resolution spectra the most basic measurement that we
can make is the iron abundance ([Fe/H]) in the atmospheres of these
RGB stars.  Iron is the typical element used to define the fundamental
metallicity of a system. From high resolution spectra we typically
have nearly 70 lines of Fe~I and around 15 of Fe~II which allows
the determination of an exceptionally secure value for [Fe/H]
independent of the usual caveats and uncertainties accompanying the
determination of abundance from a single line, and especially all the
additional problems of measuring a different element (such as Ca or
Mg) and converting this to [Fe/H].  From high resolution spectra there
are also a host of other elements for which an accurate abundance can
be obtained. These give us additional insights into the details of
past star formation processes in these galaxies (cf. McWilliam 1997,
and references therein).

Light Elements ({\it e.g.}, O, Na, Mg, Al) allow us to trace
``deep-mixing'' abundance patterns in RGB stars. This is a very
distinctive pattern of abundances that are markedly different in
globular cluster giants and field stars.  There are also the group of
$\alpha$-elements ({\it e.g.}, O, Mg, Si, Ca, Ti), elements with even
Z. The production of these elements is dominated by Type~II Supernovae
(SNII).  They affect the age estimates based on RGB isochrones, as
lower $\alpha$-tracks are bluer than high $\alpha$-tracks. The
$\alpha$-abundance limits the number of SNII explosions that can have
polluted the gas from which the star was made, and thus contains
estimates of the fraction of lost ejecta and/or IMF variations in the
stellar populations of a galaxy through time.  The Fe-Peak elements
({\it e.g.}, V, Cr, Mn, Co, Ni, Cu, Zn) are mostly believed to be the
products of explosive nucleosynthesis.  The level of the Fe-peak can
(in principle) limit the most massive progenitor that can have
exploded in the galaxy ({\it e.g.}, Woosley \& Weaver 1995).  Cu
has been thought to be primarily produced in Type~Ia Supernovae
(SNIa), and hence [Cu/$\alpha$] could be an indicator of the ratio
SNIa/SNII.  However, there is no physical basis for this
assumption. It is also possible to account for Cu evolution in the
halo using metallicity dependent SNII yields ({\it e.g.}, Timmes,
Woosley \& Weaver 1995).  Heavy Metals (Z$>$30, {\it e.g.}, Y, Ba, Ce,
Sm, Eu) enable a distinction to be made between the fraction of
s-process and r-process elements in a star, and thus again puts
detailed constraints on the number and type of past SN explosions.
The [Ba/Eu] ratio can be considered an indicator of the contribution
of AGB stars to the chemical evolution process. One has to be a bit
careful since some stars can self pollute their [Ba/Eu] ratio (but
usually they stand out as having very large Ba/Fe ratios).  Since AGB
stars have a several Gyr timescale for chemical contamination of the
ISM they provide yet another type of clock.

By looking at dSph we are studying an environment which is metal poor
([Fe/H]$<-1.0$) and often with complex SFH. This describes the
properties of the galaxies found to dominate deep redshift surveys and
hence the Universe ({\it e.g.}, Madau, Pozzetti \& Dickenson 1998).
Indeed if the standard models of galaxy formation are to be believed,
small dwarf galaxies are the oldest structures in the Universe ({\it
e.g.}, White \& Rees 1978; Moore {\it et al.} 1999; Klypin {\it et
al.} 1999).  Because dSph stars are relatively easy to age date from a
CMD if (at least) [Fe/H] is known (more accuracy is obtained if
[$\alpha$/Fe] is also known), and given that dSph {\it always} contain
ancient stellar populations, there are potentially stars carrying the
nucleosynthetic signature of the earliest epochs of star formation in
the Universe ({\it e.g.}, Abel, Bryan \& Norman 2002; Bromm, Coppi \&
Larson 2002; Heger \& Woosley 2002).  We make a comparison with the
abundance patterns we observed in the oldest stars in the dSph with
theoretical predictions of the formation of the first star(s).

\section{Observations}

We observed 15 individual RGB stars in 4 nearby southern dSph galaxies
which we can add to the 17 RGB stars previously observed in 3 northen
dSph with Keck/HIRES by SCS01 (see Table~1).  We selected RGB stars to
cover as much range in color as possible in each galaxy to sample the
diversity in [Fe/H] and age and thus gain an insight into how stellar
abundances have varied over the history of the galaxies ($\sim1-$15~Gyr).

Our observations were obtained in visitor-mode, with UT2/UVES (Dekker
{\it et al.}  2000) on two separate visitor runs on Paranal, one in
August 2000, and one in January 2001.  We used the red arm of UVES,
CD\#3, centered at 580nm, which results in a wavelength range of
480-680nm.  We obtained a resolution of $\sim$40~000 and average
S/N$\sim$30 per pixel.  The integrations varied from 2$-$4~hours
depending upon the brightness of the target and sky conditions. The
details of the individual observations and the data reduction are
contained in Shetrone {\it et al.} 2002 [hereafter, Paper~I].

Covering the wavelength range 480-680~nm we detected a range of
elements, including Fe~I, Fe~II, O~I, Na~I, Mg~I, Ca~I, Sc~II, Ti~I,
Ti~II, Cr~I, Ni~I, Y~II, Ba~II, Eu~II, La~II, Zn~II, Cu~I, Mn~I (see
Paper~I for complete tables).  This allowed us to achieve the 
comprehensive abundance analysis described in Paper~I.  We are able to
make especially reliable abundance measurements because we determine
abundances from {\it more than one line} for several elements, most
crucially Fe~II. The error analysis in abundance determinations from
high resolution spectra is complex and difficult, and the procedures
and checks we carried out are described in detail in Paper~I.

\section{Results}

\subsection{Calibration Globular Clusters}

We observed 4 previously studied stars in 3 globular clusters to
verify our data reduction techniques and confirm that our abundances
are consistent with previous measurements of the same stars with
different instruments and telescopes (see Paper~I).  We have also compiled
from the literature observations of other Galactic globular clusters
(from SCS01), and also LMC star clusters (from H00; Hill \& Spite
2001; Hill {\it et al.}, in prep) to compare the elemental abundance
patterns typical in globular clusters with those found in dSphs.

The LMC clusters are in many ways the most reliable measures of
abundance variations with time due to galaxy scale evolution because
of the relative ease to age date globular cluster stars and the spread
of ages in LMC clusters.  The data set of H00 is a useful comparison
with our results.  Care should be taken in understanding the
uncertainties in a few specific cases (see H00).  Also direct
comparisons between globular cluster and field star age-metallicity
relationships are a little uncertain, because it is well known that
the properties of star clusters do not always transfer readily to the
field star population (e.g., Shetrone 1996).  Nevertheless, the LMC
star clusters provide an age-metallicity range with which to compare
our results. Recent CaT measurements of a large sample of field stars
in the LMC (Cole et al., in prep) suggest that the LMC field star
population follows the same age-metallicity relationship as the
clusters (without the gaps).

\subsection{The Dwarf Galaxies}

For each of the four dSphs we used imaging taken at the NTT (Tolstoy
{\it et al.} in prep), and supplemented by VLT archival images to
select and determine the color of individual stars as targets for
spectroscopic observation, and where possible we found radial velocity
information about our targets from previous studies, which helps to
a priori determine membership of stars in the dwarf galaxies.

From our UVES spectra we obtain an accurate [Fe/H] measurement for
each star which allows us to compare its color and [Fe/H]
with the appropriate isochrone set to determine the age.  The previous
age determinations for the stars in Sculptor and Fornax from the CaT
observations of T01 used Padua isochrones (Bertelli {\it et al.}
1994), which assume $\alpha$-enhanced abundances ($[\alpha/Fe] =
+$0.5), and as our high-resolution measurements suggest that in the
mean dSph stars have roughly solar $\alpha$-abundance we have redone
the CaT age determinations using the new Yale-Yonsei isochrones (Yi et
al. 2001), which assume solar $\alpha$-abundance. This results in a
small shift to older ages on average, but the trend (such as it is)
doesn't seem to change markedly compared to T01.  For our UVES spectra
we have used exclusively the new Yale-Yonsei isochrones of Yi {\it et
al.} to age date the stars observed (Table~2). Comparisons were made
with Bergbusch \& vandenBerg (2001) and Bertelli {\it et al.} (1994),
see also Tolstoy (2002), and although there were differences no other
set of tracks was significantly better (or worse) than another. We
chose Yale-Yonsei because they went down to very young ages (unlike
Bergbusch \& vandenBerg) and they covered a range of $\alpha$ values
down to zero.  We also used the Yale-Yonsei isochrones to determine
the ages of the dSph stars observed by SCS01 (see Table~3).

There is a marked degree of uncertainty in the ages determined from
the RGB tracks, and an additional uncertainty coming from the errors
on the measurement of [Fe/H].  There are some stars which do not fit
{\it any} of the isochrones of the appropriate (observed) [Fe/H]
for a star. Typically they are too red (or too old) for any of the
tracks, but sometimes too blue. This makes it clear that a simple
global offset ({\it e.g.}, absolute photometric offset, reddening)
doesn't solve the problem.  It might be that there is more
differential reddening in some of these galaxies than is realized.  It
is also possible that some of the stars which are on the blue side of
RGB are in fact unidentified AGB stars ({\it e.g.} in Sculptor), and
the stars on the red side of the isochrones are weak CH stars ({\it
e.g.} Shetrone, C\^{o}t\'{e} \& Stetson 2001 found large numbers in
Draco), which means their age cannot be determined from RGB
isochrones.  This might explain the large number of relatively young
stars that we appear to identify in Sculptor from the CaT
measurements.  A basic look at the previous CMD analysis suggests that
the majority of the star formation in this galaxy should have occurred
more than 8$-$10~Gyr ago, and since the stars chosen here for
metallicity measurements were randomly selected this ought to be
reflected in the distribution of ages.  Our sample of measured
metallicities is still rather small.  An example of the problems we
have encountered in Sculptor and Fornax are plotted in
Figure~\ref{sclprob} and Figure~\ref{fnxprob}. There weren't such
severe problems in determining the ages of stars in the Carina and
Leo~I stars, as they fall into the expected color range of the
appropriate theoretical stellar isochrones, but given the problems in
Fornax and Sculptor all the isochrone ages should still be viewed with
some caution.

We checked the Galactic Na~I interstellar absorption line in all our
spectra to try and understand how reliable the reddening estimates 
we have used might
be. There is clearly evidence for variation in line width between all
the galaxies and also between individual stars in each galaxy.  These
variations could be interpreted as evidence for differential
reddening, but the correlation between Na~I line width and E(B-V) is a
very uncertain and complex problem (e.g., Munari \& Zwitter 1997; Andrews,
Meyer \& Lauroesch 2001).  The variations we find fall broadly into
the scatter of this relationship.

Looking at where our spectroscopic targets lie in a CMD and comparing
them to globular cluster RGB fiducials (Da~Costa \& Armandroff 1990)
we found that globally speaking the more metal rich stars are to the
red of the more metal poor, consistent with what you would expect from
a population where the older stars have the lower metallicity.
However, going from a determination of relative ages to absolute ages
requires accurate isochrones.  Our problems with isochrones suggest
that perhaps there are more factors than age, [Fe/H] and [$\alpha$/Fe]
that are required to uniquely define the correct isochrones for stars
in dSph. This might be due to differences in stellar evolution in
globular clusters from galaxies ({\it e.g.}, different mixing lengths,
rotation rates, binary fractions, etc), or perhaps more simply the
theoretical stellar evolution models are not able to take into account
the different abundance patterns in non-globular cluster stars ({\it
e.g.}, $\alpha$-elements, and also the light elements).  For example,
Na is (along with K) by far the most important e$^-$ donor in the
atmosphere of an RGB star, and [Na/Fe] is depleted in dSph stars
relative to stars in most globular clusters.  It seems that all
isochrones have difficulty determining the ages of stars in galaxies,
but this is not a problem in globular clusters 
({\it e.g.}, Tolstoy~2002).

There are main sequence turnoff star formation histories for all these
galaxies, so we know which age range to expect, but it is clear that
the only way to resolve these problems with isochrones in galaxies
fully is to get more spectra and see if a pattern emerges.  Not only
in understanding the star formation processes at the formation epoch,
but also in understanding the problems in matching low metallicity
stellar evolution tracks to stars in these galaxies.

The previous CaT results from T01 are plotted in the following
sections as small pink circles, and a
representative error bar is also shown on the plot.  No individual
error bars are plotted for the CaT values, because they are too large,
and make the plot unreadable (see T01).  The new UVES results are
plotted as symbols with individual error bars. The error bars are
determined from the equivalent width measurement errors in the UVES
spectra and the resulting uncertainty of this error on the age
determination based on the range of isochrones within the [Fe/H] error
bars.  We also plot for comparison other observations of age and
[Fe/H] for stars in the Galactic disc from E93 as small black
dots, and the UVES study of the LMC clusters from H00 as light blue
star symbols.

The dashed line(s) in the metallicity evolution figures describe the
most simple (closed-box) chemical evolution model (as first described
by Searle \& Sargent 1972).  Searle \& Sargent define a yield as the
ratio of the rate at which an element is produced by nucleosynthesis
and ejected into the ISM to the rate at which hydrogen is removed from
the interstellar gas by star formation. We assume that these dSph
galaxies have used up all their gas forming stars at the time of the
last star formation in the galaxy.  The [Fe/H] they achieved in
doing so is the metallicity of the most recent star formation. This
allows use to determine the value of the yield as defined by Searle \&
Sargent for each galaxy, and assuming it is constant we can work back
to how [Fe/H] has varied over time by assuming that the
star/gas ratio has been changing at the star formation rate determined
from CMD analysis. This is ad hoc, but it provides a basic estimate
of how [Fe/H] may have varied as a result of the SFH of each dSph.

We also plot in each metallicity evolution figure as a dotted line the
LMC age-metallicity relation as determined by the bursting model of
Pagel \& Tautvai\v{s}ien\.e (1998).

The Galactic disk age-metallicity observations are included only as a
broad based comparison, they have of course been differently
determined coming as they do predominantly from spectra of main
sequence turnoff stars. These have different errors and uncertainties
from the RGB spectra we use. Reassuringly, abundance measurements of
giants and subgiants in Galactic globular NGC~6397 (Castilho et
al. 2000) find similar results to a study of main sequence turnoff
stars in the same cluster (Th\'{e}venin {\it et al.} 2001).  The
Galactic disk abundances are a usefully large set of age-metallicity
evolutionary measurements with which we can directly compare our
evolutionary results.

\subsection{Sculptor Dwarf Spheroidal}

From the NTT CMD (Figure~\ref{sclcmd}) we selected stars bright enough
to be observed by UVES, and where possible we found radial velocity
information about our targets from previous studies (Queloz, Dubath \&
Pasquini 1995; Schweitzer {\it et al.} 1995).  The UVES targets, and the
stars previously observed with FORS1 to measure the CaT metallicity
are marked on Figure~\ref{sclcmd}. 

\subsubsection{Star-Formation History}

The first attempt to piece together an accurate star-formation history
determined from a CMD reaching down to globular cluster age main
sequence turnoffs was made by Da~Costa~(1984). He noted that the
intrinsic width of the RGB was probably caused by a 0.5~dex spread in
abundance (from [Fe/H] = $-$2.1 to $-$1.6), and a population of
``blue-stragglers'' (in globular cluster terminology), which could be
interpreted as main sequence turn-off stars as young as 5~Gyr old.
Kaluzny {\it et al.} (1995) made a careful study of the central region of
Sculptor and found no main sequence stars (m$_V < 21$), ruling out any
star formation over the last 2~Gyrs.  Using WFPC2, Monkiewicz {\it et
al.} (1999) have made the deepest CMD of Sculptor to date, although
they cover a tiny fraction of the entire galaxy. Their accurate
photometry in this region allowed them to conclude that the mean age
of Sculptor is similar to that of a globular cluster, but that there
was probably a spread in age during this epoch of at least 4~Gyr.

Sculptor is known to contain a small number (8) of intermediate age
carbon stars ({\it e.g.}, Frogel {\it et al.} 1982; Azzopardi {\it et
al.}  1986), of which only 1 is a bona fide AGB star, the other 7 have
been identified as CH stars (Martin Groenewegen, private communication).
There is thus very little evidence for a significant metal poor
intermediate age stellar populations ({\it e.g.}, Aaronson {\it et
al.} 1984).  Sculptor shows tantalizing evidence on its Horizontal
Branch (HB) for an unusual enrichment history (Majewski {\it et al.}
1999; Hurley-Keller, Mateo \& Grebel 1999).  

Carignan {\it et al.}  (1998) found evidence for small amounts HI gas
in and around Sculptor, although this has recently been found to be
part of a large scale complex, presumably Galactic HVC and unlikely to
be associated to Sculptor (Lisa Young, private communication).

The SFH resulting from previous studies of the stellar population is
plotted in Figure~\ref{sclsfh}. Sculptor is the only predominantly old
galaxy in our sample. It has thus much in common with the northern
dwarf galaxy sample previously looked at with HIRES on Keck (SCS01).

\subsubsection{Metallicity Evolution}

From the CaT results of T01 it was suggested that the RGB stars across
Sculptor contain a mean metallicity, $<[Fe/H]> = -1.5$ with a
significant spread $\delta [Fe/H] = 0.9$, larger than that predicted
from photometry $<[Fe/H]> = -1.8 \pm 0.3$ (Kaluzny et al. 1995; Da
Costa 1984).  The histogram plot shows a fairly sharp cut off in the
upper [Fe/H] boundary for Sculptor, with a shallow tail extending
to low [Fe/H].  The highest metallicity observed was [Fe/H]$= -
1.3$, and there are a few objects around [Fe/H]$ = -$2.1, but the
majority are clustered between [Fe/H]$ = -$2.0 and $-$1.3.  There is
no evidence for any spatial effect in the distribution of stars of
different [Fe/H].  The dispersion is too large to be consistent with
all the stars being similarly old. H400 is lying in a position in the
CMD (see Figure~\ref{sclcmd}) which might cause concern as to whether
it is an RGB or an AGB star.

The CaT results, along with the new UVES results and the comparison
observations described in $\S$3.2 are plotted in
Figure~\ref{sclzfh}. There is broadly speaking an agreement between
the UVES results and the CaT results.  The observations suggest that
Sculptor has had quite a similar metallicity evolution to that found
for the old LMC star clusters of H00 in its earliest times, but not
evolving so rapidly between 10 and 15 Gyr ago, consistent with very
little star formation in Sculptor after this time. At 4$-$5~Gyr ago
the star formation in Sculptor appears to have stopped altogether, and
contrarily increased rapidly in the LMC and this is the period of
largest deviation in the comparison between the two age-metallicity
relationships.

The evolution of [Fe/H] with age shown in Figure~\ref{sclzfh} is quite
consistent with the closed-box chemical evolution scenario also
plotted (and explained in $\S$3.2).  The largest uncertainties are the
earliest enrichment at the epoch of the first star formation in this
galaxy, where our data suggest quite a large spread in [Fe/H].

\subsection{Fornax Dwarf Spheroidal}

From the NTT CMD (Figure~\ref{fnxcmd}) we selected stars bright enough
to be observed by UVES, and we supplemented this list with targets
from a previous spectroscopic study to determine radial velocities of
stars in Fornax (Mateo {\it et al.} 1991).  The UVES targets, and the
stars previously observed with FORS1 to measure the CaT metallicity
are marked on Figure~\ref{fnxcmd}. 

\subsubsection{Star-Formation History}

Fornax is larger and more metal rich, in the mean, than the other
galaxies in our sample. It also has (unusually for dSph galaxies, and
uniquely in our sample) globular clusters.  The extended sequence of
main sequence turnoffs in the Fornax CMD indicates a long history of
star formation ({\it e.g.}, Beauchamp {\it et al.} 1995; Stetson,
Hesser \& Smecker-Hane 1998; Buonanno {\it et al.} 1999), which has
only recently ceased.  The luminosity of the brightest blue stars show
that Fornax cannot contain any stars younger than 100~Myr (Stetson
{\it et al.} 1998).  The numerous ($\sim$ 120) carbon stars with a
significant age spread in the range 2-8 Gyr ago ({\it e.g.}, Azzopardi
{\it et al.} 1999; Aaronson \& Mould 1980, 1985) testify to extensive
intermediate age star formation.  This is supported by a
well-populated intermediate-age subgiant branch and a red clump, which
require a significant population with an age of 2-4~Gyr.  Most
recently, an HST study sampling the main-sequence turnoffs of the
intermediate-age and old populations in the center of Fornax was
carried out by Buonanno {\it et al.} (1999), and they found evidence
for a highly variable SFH starting at the epoch of globular cluster
formation ($\sim$ 15 Gyr ago) and continuing until 0.5 Gyr ago.  An
ancient population is present as demonstrated by detection of a red
Horizontal Branch, slightly fainter than the Red Clump (Buonanno {\it
et al.} 1999), and of RR Lyrae variables (Stetson {\it et al.} 1998).
There is also a weak, blue Horizontal Branch so Fornax clearly must
contain only a small old, metal-poor component (despite having a
globular cluster population).

Young~(1999) looked for neutral
hydrogen in and around Fornax and found none, to the column density
limit of $4 \times 10^{18} cm^{-2}$ in the galaxy center, and $10^{19}
cm^{-2}$ out to the tidal radius.

The SFH resulting from these previous studies of the stellar
population is plotted in Figure~\ref{fnxsfh}.  Fornax appears to have
been forming stars until 2~Gyr ago, and to be dominated by an
intermediate age (4$-$7~Gyr old) population, and has evidence for only
a small number of globular cluster age stars in its field population.

\subsubsection{Metallicity Evolution}

From the CaT results of T01, the RGB stars selected across Fornax
contain a mean metallicity, $<[Fe/H]> = -1.0$ and a significant spread
of $\delta [Fe/H] = 1.0$, which is slightly larger than that predicted
from photometry $<[Fe/H]> = -1.3 \pm 0.6$ (Beauchamp {\it et al.}
1995; Buonanno {\it et al.} 1985).  In contrast to Sculptor, Fornax
has more of a sharp cut-off in the [Fe/H] distribution at low [Fe/H],
with a tail of values going out to higher values.  This suggests
different evolutionary influences.  The highest [Fe/H] observed
is nearly solar and there are few objects at [Fe/H]$ < -$1.4, with the
majority clustered between [Fe/H]$ = -$0.9 and $-$1.3, with no
evidence for any spatial effect in the distribution of stars of
different [Fe/H].

The CaT results, along with the new UVES results and the comparison
observations described in $\S$3.2 are plotted in
Figure~\ref{fnxzfh}. There is broadly speaking an agreement between
the UVES results and the CaT results.  Fornax looks like it has had
quite a similar metallicity evolution to that found for the LMC star
clusters of H00. Of course with so few measurements, in both samples,
it is difficult to say how close the similarities are. It looks like
our CaT data at the oldest epochs contain a larger spread of [Fe/H]
than is seen in the LMC sample.  Perhaps there should be more concern
in the Fornax RGB for contamination by weak CH stars, given the large
number of identified carbon stars in the galaxy. This might result in
an over estimate of the ages of the oldest stars, and explain the
rather high [Fe/H] (apparently) observed in the oldest stars (in both
CaT and UVES results).  From the CMD analysis it is not expected that
Fornax should have a very high star formation rate at oldest times,
which makes it surprising that we apparently see quite a few old stars
in the CaT sample. This is consistent with the hypothesis that we may
have contamination of the RGB with weak CH stars.

The closed box chemical evolution model plotted on Figure~\ref{fnxzfh}
shows reasonable agreement with the [Fe/H] measurements going aback to
about 10~Gyr ago. Before this period, during the epoch of earliest
star formation in Fornax [Fe/H] is somewhat higher than might be
expected. This could be due to the inherent uncertainty in determining
the star formation rates at the oldest times. It resembles the G-dwarf
problem know in our Galaxy (e.g., Pagel 1997) - there is an apparent
lack of low [Fe/H] stars in Fornax, which is hard to explain within a
simple enrichment model, starting from a zero metallicity
proto-galaxy.

\subsection{Carina Dwarf Spheroidal}

From our NTT CMD we selected stars bright enough to be observed by
UVES, and we supplemented this list with targets from a previous
spectroscopic study to determine radial velocities of stars in Carina
(Mateo {\it et al.} 1993). These stars had the advantage of known radial
velocities.  The UVES targets are marked on Figure~\ref{carcmd} which
is a VLT/FORS1 CMD made from archival data. 

\subsubsection{Star-Formation History}

Carina was the first dSph galaxy to benefit from deep imaging in the
early days of modern CCD detectors (Mould \& Aaronson 1983) and
remains one of the more spectacular examples of a dSph galaxy with
significant variations in star formation rate with time, and a
dominant intermediate-age population (Mighell 1990, 1997; Smecker-Hane
{\it et al.} 1994; Hurley-Keller, Mateo, \& Nemec 1998; Hernandez et
al. 2000; Dolphin 2002).  The bulk of the stellar population in Carina
has ages ranging from about 4 to 7 Gyr, though older (RR~Lyr
variables: Saha, Seitzer, \& Monet 1986) and younger stars (or blue
stragglers: Hurley-Keller {\it et al.}) are also clearly present.

The most recent quantitative star-formation histories for Carina (from
two different data sets) are plotted in Figure~\ref{carsfh}. There are
obvious differences between the three different SFHs.  It is hard to
normalise the absolute star formation rates between the three
different determinations, so this has been done in an arbitrary
fashion. The main difference is that Hurley-Keller {\it et al.} (1998)
in contrast to Hernandez {\it et al.} (2000) and Dolphin (2002) see
discrete bursts of star formation. This might be because Hurley-Keller
{\it et al.} are better able to resolve such details as they cover a
much larger area of the galaxy and hence all the main sequence
turnoffs are better populated than in the tiny HST field used by the
other two studies.  Also Dolphin and Hurley-Keller {\it et al.} find
evidence of ancient star-formation (as expected from the RR~Lyr
population found by Saha {\it et al.} 1986), whereas Hernandez {\it et
al.} do not.  Another difference is the assumed [Fe/H], not only the
absolute value, but the spread. Dolphin finds [Fe/H]=$-1.2\pm$0.4,
where as Hurley-Keller {\it et al.} assumes [Fe/H]=$-2.1\pm$0.1 and
Hernandez et al. [Fe/H]=$-2.0\pm$0.2. In comparison with our results
here it looks like Dolphin is a little too metal rich, and the other
two studies a little too metal poor. These different assumptions
significantly affect the star formation history determinations, and
particularly the age of the peak(s) in the star formation rate.

\subsubsection{Metallicity Evolution}

The photometry of Carina from wide field imaging reveals a very thin
giant branch ({\it e.g.}, Hurley-Keller {\it et al.} 1998).  Given the
large age spread measured from Main sequence turnoff photometry in
this galaxy, a metallicity spread has always been thought necessary to
compensate and produce the observed narrow RGB.  Our UVES results for
5 stars give an average abundance, $<[Fe/H]> = -1.6$, with a spread
$\delta [Fe/H] = 0.5$.  This is different from previous determinations
from CMDs, which gave [Fe/H]$ = -2.0$, and a spread $\delta[Fe/H] <
0.1$ (Mould \& Aaronson 1983; Smecker-Hane {\it et al.} 1994;
Hurley-Keller {\it et al.} 1998).  CaT spectroscopy of 15 giants in
Carina (Da~Costa 1994) resulted in a narrow average [Fe/H]= $-1.9\pm
0.1$, with one giant more metal-poor ([Fe/H] = $-$2.2). This is
significantly more metal poor than our UVES results, but similar to
our results they found a very small spread in [Fe/H].

We plotted the [Fe/H] versus age measurements as blue star symbols in
Figure~\ref{carzfh}.  The several (different) SFHs (shown in
Figure~\ref{carsfh}) are represented by a series of lines, all
consistent with a very small spread in [Fe/H]. There is no resolvable
difference between the three predictions. They are broadly speaking in
agreement with the UVES results, with one outlier (M10) at lower
[Fe/H] than might be expected. This star is comparable to the previous
CaT determinations, and may represent a small population at this
metallicity (similar to the old LMC cluster metallicity) that we don't
sample very well.

Carina is the faintest galaxy in our sample, although its mass and
average [Fe/H] are similar to Sculptor (see Table~1).  There is only
evidence for fairly mild evolution (very small spread) in [Fe/H], but
the number of stars observed is still very small.  Unlike Sculptor and
Fornax, in Carina the metallicity evolution looks significantly
different to the LMC measurements of H00. Carina has a metallicity
evolution which is consistently lower than the LMC clusters.  The one
star (M10) which is more metal poor than might be expected from the
amount of star formation creates a spread in [Fe/H] at the oldest
ages.  The [Fe/H] values from the independently determined SFHs
suggest that the older stars in Carina should have [Fe/H] similar to
the older LMC clusters. However, our results suggest (as do those of
Da Costa 1994) that at least a small fraction of stars in Carina are
of very low [Fe/H]. This may represent the ``shot-noise'' in the self
enrichment of this galaxy at the earliest epochs of star formation, or
it might be a result of the highly variable SFH in Carina.  The number
of high resolution observations we have is still too small to quantify
this accurately, but the hints are tantalizing (see also Paper~I).

\subsection{Leo~I Dwarf Spheroidal}

From the previous spectroscopic study of Leo~I (Mateo {\it et al.}
1998) we selected stars bright enough to be observed by UVES. These
stars had the advantage of known radial velocities.  The UVES targets
are marked on Figure~\ref{leocmd} which is a VLT/FORS1 CMD made from
archival data. 

\subsubsection{Star-Formation History}

Leo~I is the most distant galaxy in our sample and this along with the
proximity on the sky of the bright Galactic star Regulus have made
photometric studies difficult.  The earliest observations of Leo~I
indicated a significant intermediate age stellar population: Hodge \&
Wright (1978) observed an unusually large number of anomalous
Cepheids; carbon stars were found by Aaronson, Olszewski \& Hodge
(1983) and Azzopardi, Lequeux, \& Westerlund (1985, 1986).  High
quality CMDs have recently revealed a horizontal branch well populated
from blue to red in Leo~I (Held {\it et al.} 2000), and subsequent
observations were made of RR~Lyr variable stars (Held {\it et al.}
2001) predicted from the detection of the Horizontal Branch
population.  Held {\it et al.} estimated a mean metallicity for Leo~I
RR~Lyr stars of [Fe/H]=$ -1.8$, using the relation derived by
Sandage~(1993) between the average period of the RRab variable stars
in the Milky Way globular clusters and their metallicity.  This might
suggest that the oldest populations in Leo~I ($>$10~Gyr) have a mean
metallicity close to this value.

Sophisticated modeling of main sequence turnoffs in a deep HST CMD by
Gallart {\it et al.} (1999) has given the most accurate information on
the SFH of Leo~I to date. The SFH they derived is plotted in
Figure~\ref{leosfh}. Also plotted is a more recent determination by
Dolphin (2002) based on the same data set.  The Gallart {\it et al.}
analysis suggests a smoothly varying metallicity evolution of the
stellar population between [Fe/H]$= -1.4$ and [Fe/H]$= -2.3$
(consistent with the mean determined by Held {\it et al.}).  Dolphin
on the other hand found much higher metallicities, [Fe/H] between
$-0.8$ and $-1.2$ dex. This discrepancy shows the inherent
uncertainties in determining [Fe/H] from the color of the RGB
for a galaxy with a significant intermediate or young stellar
population. Surprisingly the SFHs look very similar from both
studies. Similarly to the case of Carina our high resolution [Fe/H]
measurements seem to fall somewhat intermediate to the two photometric
determinations.

\subsubsection{Metallicity Evolution}

The Leo~I CMD reveals an RGB which, based on photometry, can (and has
been) suggested to be either indicative of a broad range of
metallicity in the galaxy ([Fe/H]$=-1.9\pm0.5$ dex; Gallart {\it et
al.} 1999) or a narrower range of higher metallicity stars
([Fe/H]$=-1.\pm0.2$ dex; Dolphin 2002).  Our UVES results are only for
2 stars, which makes any statistical analysis difficult. The
abundances are [Fe/H]$=-1.06\pm0.2$ and [Fe/H]$=-1.52\pm0.2$, which
falls somewhat intermediate between the two CMD analyses, but with
only two stars it is impossible to be conclusive. Clearly more
spectroscopic data is needed for this to be confirmed.

We plotted the [Fe/H] versus age measurements as blue stars in
Figure~\ref{leozfh}.  There is no resolvable difference between the
two chemical evolution scenarios, and they both, broadly speaking,
agree with the UVES results, although the (young) high [Fe/H]
point is slightly above what might be expected.

It is very difficult to say anything detailed about the metallicity
evolution in Leo~I with just two measurements.  In comparison to the
trend of the LMC star clusters from H00, there appears to have been no
rise in enrichment levels at intermediate ages ($\sim$8~Gyr ago),
consistent with a galaxy dominated by star formation occurring in the
last 5~Gyr (Figure~\ref{leosfh}).  The metallicity evolution in Leo~I
may be similar to that observed in Carina, with very little spread
until quite recent times.

\newpage
\section{Other Elements}

There is much more information in a high resolution spectrum than the
iron abundance of a star. There are a host of lines from other
elements which can be interpreted in terms of how the chemical state
of the interstellar medium from which stars are made has varied over
time.  There are four broad categories into which elements are
distributed depending upon their properties: Light Elements,
$\alpha$-Elements, Iron-Peak Elements and Heavy Elements. All provide
pieces of the overall story of the chemical evolution history of the
stellar populations in a galaxy.

Our sample of dSph galaxies has a very broad range of ages of stars,
and this makes studies of the abundance variations with time possible.
Most of the diverse (and complex) aspects involved in interpreting
these kinds of measurements are covered Paper~I.  However here we
would like to summarise those aspects we consider most crucial for
gaining a clearer understanding of the global star formation
properties of these galaxies and how they may have formed and evolved
through time.

\subsection{The $\alpha$-Elements}

The $\alpha$-elements is a common collective term for some even-Z
elements (e.g., O, Mg, Si, S, Ca and Ti), that are predominantly
dispersed into the ISM in SNII explosions.  They are typically over
abundant in globular clusters and halo stars relative to (solar) disk
stars. The traditional interpretation of the trend of [$\alpha$/Fe]
decreasing with time is that this is due to the time delay between
SNII and SNIa ({\it e.g.}, Tinsley 1979; Gilmore \& Wyse 1991).  In
principle the mixture of different $\alpha$-abundances can provide a
clue as to the mixture of SN explosions that must have produced the
enrichment seen in a star ({\it e.g.}, Woosley \& Weaver 1995),
because different mass SN produce different proportions of
$\alpha$-elements.

We define an average $\alpha$ in the same way as SCS01, namely, as an
average of Mg, Ca and Ti abundances ([$\alpha$/Fe] =
1/3([Mg/Fe]+[Ca/Fe]+[Ti/Fe]). There are different ways to define this
since O, Mg, and Si dominate by mass and relative abundance they can
also dominate the $\alpha$-effects on chemical evolution models and
isochrones.  In Figure~\ref{alf1} we compare the distribution of
$\alpha$-elements in our data (the red and blue symbols as defined in
the caption) and the Keck measurements made by SCS01 of northern dSph
RGB stars (Draco, Ursa Minor and Sextans), as green triangles with
similar measurements of disk stars (E93, black crosses); halo stars
(McWilliam {\it et al.} 1995, open squares); Galactic globular cluster
measurements (this work and SCS01, blue crosses); LMC star cluster
measurements (H00, blue stars).  SCS01 did not work out ages for their
dSph stars, and so we have made estimates based on the [Fe/H] and
color of the stars they observed (see Table~3).

We have plotted the $\alpha$-abundances both in the ``traditional''
manner, against [Fe/H] in the upper panel of Figure~\ref{alf1}, and
against age in the lower panel. Both provide differing insights as to
how these galaxies may be evolving with time, and also how our
observations compare with those of other galaxies. Plotting against
age is more useful from the point of view of understanding chemical
evolution, but it is not easy to find suitable measurements to compare
our results with, as it is difficult to determine accurate ages for
halo or bulge stars.  

The dSph $\alpha$-abundances when plotted against age in the lower
panel of Figure~\ref{alf1} appear to follow the same distribution as
those for the disk and the star clusters.  However, if we look at the
plot of [$\alpha$/Fe] versus [Fe/H] in the upper panel of
Figure~\ref{alf1}, the properties of dSph are clearly significantly
different from the disk in the sense that although the levels and the
variation with stellar age of [$\alpha$/Fe] are similar between the
disk and the dSph, this is all occurring at much lower [Fe/H] in the
dSph.  The upper panel of Figure~\ref{alf1} also shows that the
properties of the dSph differ from those typical of halo stars,
although there is more overlap in [Fe/H] values, the [$\alpha$/Fe] of
the halo stars are typically higher, although there is a population of
low [$\alpha$/Fe] halo stars which could conceivably be accreted
satellite galaxies (e.g., Nissen \& Schuster 1997). The star cluster
measurements follow the upper envelope of dSph values, although most
of the cluster stars show deep mixing abundance pattern. Thus the
properties of [$\alpha$/Fe] in dSph does not match globular clusters,
nor the disk nor the majority of the halo.

\subsubsection{The Evolution of [$\alpha$/Fe] due to star formation}

In Figure~\ref{alf5} we plot [$\alpha$/Fe] for each dSph separately,
over-plotted is an illustrative {\it estimate} of the variation of
[$\alpha$/Fe] for each galaxy {\it given the star formation rate
variation}. There really are not sufficient data on these galaxies to
be certain that we are seeing direct evidence of evolution in
[$\alpha$/Fe], but the results are highly suggestive. The dashed lines
are not derived from the SFH directly, but a knowledge of the SFH is
used to find the most likely pattern with time in the
$\alpha$-abundances.  Carina has the most impressive evidence for
evolution of abundances due to variations in star formation rates with
time.  The variations seen in [$\alpha$/Fe] are supported by
consistent variations in Ba, La, Nd and Eu (see Paper~I).  With the
exception of Carina, the star formation histories {\it inferred} from
[$\alpha$/Fe] are consistent with the CMD SFH described in $\S$3. In
Carina it is possible that due to the incorrect metallicity evolution
assumed in all the CMD SFH determinations described in $\S$3.5.1 that
the SFH has not been well determined. Figure~\ref{alf5} suggests that
there has been an ancient epoch of star formation (13$-$15~Gyr ago),
followed by a hiatus, then another epoch of star formation 9$-$11~Gyr
ago. More data are clearly required to quantify this interpretation of
these data.

One curious point is that in all our observations [$\alpha$/Fe] is
roughly solar for the oldest stars. This might be a remnant of the
initial enrichment of the dSph gas in the early universe, by a process
which is quite different from the star formation processes that we see
today. This is certainly not consistent with what would be expected
for stars switching on in a zero (or close) metallicity environment.
It ought to take time for the [$\alpha$/Fe] to get to zero (e.g.,
Pagel \& Tautvai\v{s}ien\.e 1998).

\subsubsection{Comparison with Common ``Metallicity'' Determinants}

In Figure~\ref{alf2} we plot three elements of particular interest
against age, using the same symbol definitions in Figure~\ref{alf1}.
We choose to look at [O/Fe] against age in the upper panel because O
is so abundant, and thus arguably the best ``metallicity'' tracer.  In
middle panel of Figure~\ref{alf2} we look at [Ca/Fe] because of our
interest in accurately interpreting CaT results, The high resolution
measurement of the [Ca/Fe] abundance is a useful test of the
reliability of the conversion between CaT observations and [Fe/H].
Another common tracer of [Fe/H] from low resolution spectra is [Mg/Fe]
which we plot in the lower panel of Figure~\ref{alf2}. We also plot in
all the panels of Figure~\ref{alf2} the E93 disk abundances as black
crosses.

There is quite a large scatter in the [O/Fe], especially at the
earliest times. It follows the same trend as the disk abundances, but
the scatter is somewhat larger than seen in the disk at earlier times,
often similar to the halo enhancement.  The relatively young
($\sim$~2~Gyr old) LMC star cluster NGC~1978 appears to have a very
high oxygen abundance, but this is a very insecure measurement (see H00).

The [Mg/Fe] results look very similar to the [Ca/Fe], with slightly
greater spread in the [Mg/Fe] values. These values are also fairly
similar to the results from the disk stars.  They are very close to
solar with little or no discernible trend, although in common with
[O/Fe] the dSph measurements appear to have a larger scatter at the
oldest ages than is seen in our disk.  Looking at the 3 stars we
observed in Fornax, they appear to have a fairly constant above solar
value Ca abundance ([Ca/Fe]$\sim$0.25), and no similar enhancement in
[O/Fe] or [Mg/Fe].  There have been breaks in the $\alpha$-abundance
patterns seen before ({\it e.g.}, in the bulge, McWilliam \& Rich
1994), but these previously observed patterns are different to those
we observe in Fornax.  The reasons for these breaks are not clear.
Obviously considerable care has to be exercised in interpreting
metallicity determinations based on CaT and Mgb lines. They are
clearly statistical measurements requiring large samples of stars.

\subsubsection{Interpretation of [$\alpha$/Fe] observations}

The low [$\alpha$/Fe] values found in the disk stars of our Galaxy
have been interpreted as evidence for star-formation in material with
a large fraction of SNIa ejecta. This is perhaps not surprising for
our disk, with high metallicity, and typical predictions of fairly
recent formation (from enriched material).  It is not clear that the
same assessment can be made of the similarly low [$\alpha$/Fe] in dSph
galaxies.  Everything we know about dwarf galaxies suggests that they
have never had very high rates of star formation. The stars in dSph
typically have much lower [Fe/H] and [O/H] than in our disk. A low
star formation rate may explain the low [$\alpha$/Fe] values because
the SNII products may predominately come from low mass SNII (8$-12
\Msun$ progenitors), which result in lower $\alpha$ yields than
their higher mass cousins ({\it e.g.}, Woosley \& Weaver 1995).  This
is (unfortunately) effectively a truncated IMF, but it is motivated by
the likelihood that in the physical conditions to be found in small
galaxies the probability of forming high mass molecular clouds (and
thus high mass stars) is low.  The variations in [$\alpha$/Fe] shown
in Figure~\ref{alf5} are suggestive that [$\alpha$/Fe] does vary with
star formation rate, but more data are needed to confirm and quantify
this.

\subsection{Iron-Peak Elements}

The Iron-peak elements ({\it e.g.}, Cr, Mn, Co, Ni, Cu, Zn) are the
highly stable end-products of the nucleosynthesis sequence by nuclear
fusion, and they provide the seeds which capture neutrons and produce
heavy elements.  It is hard to find suitable comparison samples with
known ages for heavy element abundances because it seems that none of
the observations of these elements have been made of stars of known
age.  Generally speaking it seems that dSph stars have halo-like
Iron-Peak abundance patterns (see Paper~1). Two important Fe-peak
elements with regard to understanding galaxy evolution and also making
the connection between galaxies observed in the nearby and more
distant Universe are Zn and Cu.

\newpage
\subsubsection{Zinc}
Zn plays a pivotal role in interpreting the abundance patterns of the
Damped Lyman Alpha systems (DLAs), so it is interesting to compare it
with the values measured in the nearby universe.  In Figure~\ref{zncu}
[Zn/Fe] is plotted versus [Fe/H]. The dSph observations are plotted
using the same symbols defined in Figure~\ref{alf1}.  Also plotted in
Figure~\ref{zncu}, for comparison, compilations of [Zn/Fe] for disk
and halo stars obtained from the literature ({\it e.g.}, Sneden,
Gratton \& Crocker 1991), plotted as black stars, and the black open
circles are a selection from the new sample of disk and halo star
measurements from Primas {\it et al.}  (2000).  Also plotted are the
values measured in DLy$\alpha$ absorption systems at cosmological
distances by Pettini {\it et al.}~(2000), as black circles, with purple
error bars.  The DLA [Zn/Fe] measurements at high redshift are, on
average, higher than anything else observed in the Local Group (see
Figure~\ref{zncu}). 
This is largely due to depletion of Fe onto dust grains. 
In fact, these particular DLAs were selected by Pettini {\it et al.}
as those that showed the lowest dust depletions such that the 
gas phase Zn results should represent the total Zn abundance quite well.   
The fact that they are quite homogeneous could imply that the Zn 
abundances are quite uniform in these DLA systems, but of course the
uncertainties in the dust corrections complicates the interpretation.
Also, the DLA measurements are very different from all others in 
Figure~\ref{zncu}, coming from absorption of quasar light in the 
interstellar gas of an intervening galaxy.

The nucleosynthetic origin and history of Zn is not well understood
(see Paper~1).  Traditionally, Zn has been thought to be produced by
s-process neutron capture reactions, but then it should decrease with
metallicity (e.g., Matteucci {\it et al.} 1993).  Another possible
site of Zn production is explosive Si-burning in SNII. Recent
computations (Umeda \& Nomoto 2002) suggest that the over-solar
[Zn/Fe] values found in the most metal-poor stars of our Galaxy are
the result of deep mixing of complete Si-burning material and
significant fall-back in SNII.  It is clear that Zn is a key element
to further constrain the physics of SNII explosions, but more data
will be needed (also of other elements) before being able to discern
among the different possible solutions (e.g., Primas {\it et al.} in
prep., from the VLT Key Programme 165.N-0276).

\subsubsection{Copper}

In the case of Cu, it has been suggested that the most significant
production occurs in SNIa ({\it e.g.}, Matteucci {\it et al.} 1993).
Cu is clearly underabundent with respect to Fe in all our observations
of dSph (e.g., Figure~\ref{cualf}).  They resemble the halo star Fe-peak
abundance patterns, which might suggest a lack of SNIa enrichment of
the dSph ISM, but given that dSphs also show lower $\alpha$-abundances
relative to metal-poor halo stars
(Figure~\ref{alf1}) interpreted as a low SNII rate
(relative to SNIa), it is not clear that this simple explanation is
valid here. In Paper~I we suggest that Cu is not produced readily in
{\it low metallicity} SNIa.  Another possible explanation is that dSph
galaxies rarely form very massive stars (i.e. $> 15\Msun$, see
$\S$4.1).  SNII explosions at the low mass end of the possible mass
range agree with the solar [$\alpha$/Fe] measurements and also with
the apparent underproduction of heavy elements.  It is also possible
to invoke metal dependent SN yields (e.g., Timmes {\it et al.} 1995;
see discussion in Paper~I).

[Cu/$\alpha$] has been supposed to be a good indicator of the ratio of
SNIa/SNII, assuming that Cu is predominantly made in SNIa, and
$\alpha$-elements are predominantly SNII products.   Our data do
not support this though.    [Cu/$\alpha$] is very low (halo-like) in
all of the dSph stars (with the exception of the most metal rich star
in Fornax; see Figure~\ref{cualf}).   Also, [Cu/$\alpha$] is remarkably
constant with age.   This again suggests that Cu is not significantly
produced in the {\it low metallicity} SNIa, like alpha-elements,
especially given the star formation histories in these galaxies and the
expected SN Ia contributions before $<$5 Gyr when the Cu abundance
begins to increase.

\subsection{The Heavy Elements}

The Heavy elements are defined as those elements beyond the iron peak
(Z$>$31). Heavy elements can only be synthesized by repeated neutron
captures onto to Iron-peak elements followed by $\beta$ decay. This
can happen either slowly, with respect to the time for $\beta$ decay
(s-process) or rapidly (r-process), and these two processes are the
result of different neutron sources and lead to different abundance
patterns.

Ba is thought to be a predominantly s-process element, which is
thought to mainly occur in low mass ($1-3 \Msun$) AGB stars ({\it
e.g.}, Truran 1981; Busso {\it et al.} 1995; Lambert {\it et al.}
1995).  Eu is a nearly pure r-process element and thought to be
produced only in SNII. It is typically enhanced in the stars measured
in dSph, as is in our halo.  This is also true for all but the oldest
of the LMC star clusters. This is perhaps somewhat at odds with the
relatively low $\alpha$-abundances measures in the dSph stars, but
consistent with our interpretation of $\alpha$-element and Fe-peak
abundances being predominantly from low mass SNII (predominantly
8$-$12~$\Msun$) in dSph, as discussed in $\S$4.1 and $\S$4.2.

In Figure~\ref{baeu}, we have plotted the [Ba/Eu] abundances, the
symbols and the comparison data is the same as in Figure~\ref{alf1} in
$\S$4.1.  As in previous figures we plotted the same data twice, once
in the more traditional manner against [Fe/H], in the upper panel of
Figure~\ref{baeu}, and again versus age to put the data directly in an
evolutionary context, in the lower panel.  Examination of these two
figures gives us an idea of the mix of s and r-process products in the
stars in our sample, which allows us to assess the contribution of AGB
stars to the chemical evolution of dSph as compared to low mass SNII.
In the lower panel of Figure~\ref{baeu} we can see that, with the
exception of one star (in Ursa Minor, which is clearly s-process
contaminated, with a huge over abundances of the s-process elements
compared to Eu), that the older stars in our sample show {\it
predominantly} r-process heavy elements.  This is similar to the older
disk stars, and appears to evolve with time to include more and more
s-process heavy elements.  In the upper panel of Figure~\ref{baeu} we
can see that the same domination of r-process elements is present in
the halo and bulge stars, although there is some scatter. The LMC star
cluster results plotted in the upper panel of Figure~\ref{baeu} show
either a balance of r- and s-process elements, or perhaps a slight
enhancement of s-process elements. This is different from the Galactic
globular clusters and the halo stars (in the mean), and more similar
to the younger disk stars (see also the lower panel of
Figure~\ref{baeu}).

The heavy elements in the oldest stars in the Universe must come from
the r-process, because AGB stars have not yet had time to evolve and
contribute to the ISM (takes at least $\sim$1~Gyr).  In both panels of
Figure~\ref{baeu} we see a trend of increasing contribution of
s-process elements with increasing [Fe/H] (and age) across the disk,
halo and also the dSph sample, consistent with the interpretation
that it takes a few Gyr for the low mass stars created in the earliest
star formation epochs to start producing enough s-process elements to
influence subsequent generations of stars. Our data allows us to
constrain this delay in our sample of dSph galaxies until about
$\sim$10~Gyr ago. It is at this time that s-process contributions
start to match those of r-process elements, at least in Fornax,
Sculptor and Carina. It doesn't look like Leo~I ever reached this
stage, nor Draco. But as in all our conclusions, this is uncertain due
to the small number statistics here.

\section{Implications for Understanding Galaxy Evolution}

Having such detailed information, albeit on very few stars, it is
interesting to try and put our results into a context that can be
understood in terms of the implications for our understanding of
galaxy evolution.  This has been termed {\it Chemical Tagging} by
Freeman \& Bland-Hawthorn (2002), and is probably the only way to
accurately disentangle the different phases in a galaxy's evolution
back to the epoch of formation, tagged by the ever increasing
enrichment of an interstellar medium by elements produced in stars and
liberated in supernovae explosions or stellar winds.  The galaxies we
are concentrating on, dSph, are arguably very simple objects with
regard to chemical evolution. They are like a single cell star forming
entity.  They are small potentials in orbit around a much larger one
and are thus unlikely to accrete much (if any) extraneous matter
during its lifetime (either intergalactic gas, or galaxies) because
they will typically lose the competition with our Galaxy. Their
evolution is going to be influenced, maybe strongly, by the presence
of our Galaxy. They will permanently lose gas to our Galaxy if it is
expelled too far from their centers, and the dynamical friction of the
orbit may have a strong influence on the rate of star formation at any
given time in these galaxies.

\subsection{Similar Abundance Patterns: Consistent Star Formation Mode}

One aspect of these dSph RGB abundances which clearly stands out is
how uniform they appear to be in general, despite a range of different
star formation histories, and also how different they are from the
properties of stars observed in the disk and halo of our Galaxy ({\it
e.g.}, $\alpha$-abundances, Figure~\ref{alf1}).  It is as if all stars
know the mass of the potential in which they are forming. 

Figure~\ref{alf5} suggests that with our UVES results we can trace
the variation of [$\alpha$/Fe] with time caused by varying (but always
low) star formation rates at different times.  This is the first time
it has been possible to directly {\it measure} the [$\alpha$/Fe]
evolution of a stellar population over Gyr time scales, back from the
earliest epoch of formation to the most recent star formation.  The
range of variation of alpha is quite small (which means accurate
measurements are required to observe it), and it never reaches the
parameter space where the disk and halo are stars predominantly to be
found. So probably star formation, when it occurs, always occurs at
similarly low levels in these small galaxies. Perhaps this is not
surprising as it is the mass of cool HI gas in conjunction with the
ability of the gravitational potential to compress it that determines
the size and number of molecular cloud complexes which can form
stars in a galaxy. It seems very unlikely that dwarf galaxies have ever
experienced a very high star formation rate, and have thus had very
few (if any) SNII of masses more than $15-20\Msun$.

Thus it seems as if the {\it classical} interpretation of [$\alpha$/Fe] as
applied to the disk and halo of our Galaxy doesn't work when applied
to dSph. 
It is possible
to produce solar [$\alpha$/Fe] values from very low star formation
rates. Assuming a Salpeter IMF and using the predicted yields from
Woosley \& Weaver (1995), an event which forms 2500$\Msun$ of stars
(over 10$^7$yrs; sfr$\sim2.5\times 10^{-4}\Msun$/yr) reduces O 
and Mg by a small amount ($\sim0.1-0.15$ dex) and has no net
effect on Si. However, for an event which forms only 1000$\Msun$
stars (sfr$\sim10^{-4}\Msun$/yr) the affect on O, Mg and Si is
significant ($\sim0.5-0.6$ dex).  This means that a low star formation
rate can explain the low [$\alpha$/Fe] we observe in these galaxies.
It also means that [$\alpha$/Fe] will not be a good indicator of SNIa
rates in systems with few massive SNII.

\subsection{Chemical Evolution: Evidence for infall or outflow}

It is somewhat uncertain if these systems were (or are) small enough
to lose significant mass (and metals) in SN explosions. This is partly
due to problems determining the total mass of these small galaxies,
both today and ascertaining what they may have been in the past.  If
we believe current estimates ({\it e.g.}, Mateo 1998, and references
therein), these galaxies are currently of low enough mass that they
are easily in the mass-range where supernova explosions ought to
result in a significant metal-enriched blow-away, ({\it e.g.}, Mac Low
\& Ferrara 1999; Ferrara \& Tolstoy 2000; Mori, Ferrara \& Madau
2002).  Blow-away means that it is unlikely that these galaxies would
easily lose their entire interstellar medium in this fashion, but they
may suffer metal enriched winds, where the metals may leave the system
but the gas probably will not.  Even if the metal enriched winds
escape the central regions of the galaxy, it seems likely that they
will continue to exist in a bubble surrounding the galaxy ({\it e.g.},
Ferrara, Pettini \& Shchekinov 2000; Mori, Ferrara \& Madau 2002), and
therefore may still play a role in future star formation episodes.  If
dwarfs have a more extended DM halo than currently thought ({\it
e.g.}, Kleyna {\it et al.} 2002) these metal rich bubbles will be
smaller, and more likely to collapse back into the central potential.
None of the dSph around our Galaxy contain evidence for any
particularly violent or intense star formation episodes in their past,
so the typically high (simultaneous) SN rates assumed in simulations
modeling the blowout properties are unlikely to occur in dSph, {\it
e.g.}, Mori {\it et al.} assume a sfr = 0.03$\Msun$/yr, which is much
higher than is thought to be typical for small dwarfs
($\sim10^{-4}\Msun$/yr). However it is difficult to distinguish
between low yields due to a low star formation rate and low yields due
to mass loss.  Another possible explanation is that the products of
more massive ($> 20\Msun$) SNII are predominantly lost to these
galaxies, but it is hard to find any plausible reason why this might
be so.

With the exception of the earliest times all abundance properties in
these dSph are consistent with closed-box evolution as inferred from
the star formation histories (Figures~\ref{sclzfh}, \ref{fnxzfh},
\ref{carzfh} \& \ref{leozfh}).  Our data are not sufficient to rule
out a leaky box scenario where some (small) fraction of the metals are
lost.  The implication is that outflows (and inflows) of metals are
not necessary to explain the chemical evolution observed through most
of the history of any of these systems.  Gas and metals may flow out
from the galaxy ({\it e.g.}, Mac Low \& Ferrara 1999; Mori, Ferrara \&
Madau 2002) but, with the possible exception of the initial burst of
star formation, there is a chance that they will return on a timescale
($<1-2$Gyr) compatible with the evolutionary timescale of star
formation in these galaxies.

It is possible that the large scale differences we see in star
formation histories of dSph result from their differing orbits around
our Galaxy ({\it e.g.}, Oh, Lin \& Aarseth 1995).  Tidal perturbations
can significantly effect star formation, either compressing the ISM
and increasing the star formation rate, or inhibiting it by stretching
the galaxy, or even stripping the ISM altogether and thus stopping
star formation for good, independent of the star formation rate in the
galaxy.  Some galaxies apparently lost (or used up) all of their gas
very quickly (in a few Gyr), and others have managed to continue
forming stars until very recently, but all seem to maintain similar
chemical abundance patterns across time, until perhaps the last few
Gyr. This suggests that the properties of star formation have not
changed significantly over time.  There is evidence for slow evolution
in both the SNIa/SNII ratio, and the increasing contribution of
s-process elements over time.  There is no significant evidence for
large discrepancies in abundances such as might be expected from gas
renewal at some point in the life of a galaxy from a pristine source,
but more data are required to make more firm conclusions.

\subsection{Building Large Galaxies from Smaller Ones}

One of the fundamental pillars of CDM theory is that small galaxies
are the building blocks of larger ones ({\it e.g.}, Navarro, Frenk \&
White 1995). The timescale on which this build up of material occurs
has been a matter of some discussion between the classical approach to
modeling within the CDM paradigm which results in the prediction that
galactic disks and halos (such as our own) have relatively recently
experienced significant accretions of smaller galaxies ({\it e.g.},
Steinmetz \& Navarro 2002; Kauffmann, White \& Guiderdoni 1993), and
those approaches which, independent of assumptions from CDM, utilise
observations of our Galaxy and the nearby Universe to place limits on
the star formation rates for all the different components in our
Galaxy over a Hubble time ({\it e.g.}, Prantzos \& Silk 1998;
Rocha-Pinto {\it et al.} 2000).

Our results unequivocally show that the stars observed in dSph
galaxies {\it today} (many of which are extremely old) {\it cannot} be
used to make up the majority of stellar mass in our Galaxy, neither in
the disk nor in the inner-halo (nor the bulge) because their
nucleosynthetic signatures are not compatible. This places a limit to
the age at which the majority of merging of small halos can have
occurred, if it is to have occurred.  This is of course assuming that
these small halos are similar to the dwarf galaxies we see today. The
majority of these kinds of mergers must have occurred very early, in
the first few Gyr of structure formation, because this is the only way
to ensure that the large potentials accrued mostly gas and very few
stars, and so the majority of star formation will occur in a much larger
potential than a dwarf galaxy with a lot more influence from massive
SNII which is required to explain the abundance patterns seen in our
Galaxy, but not found in dSph.  In this scenario there still remains the
mismatch in the M/L between dwarf and giant galaxies; dwarf galaxies
are frequently dark matter dominated, unlike large spiral galaxies.

Of course our dSph results do not
place any limits on the effect of significantly larger accretions,
({\it e.g.}, LMC like objects).  The detailed abundance measurements
of old and intermediate age stars in the Magellanic Clouds are rather
sparse, and so the conclusions are inherently uncertain, but there are
suggestions that the abundance patterns of the stellar population
observed in the Clouds and nearby dwarf irregulars do not resemble our
Galaxy anymore than the dSph do ({\it e.g.}, Hill 1997; H00; Venn {\it
et al.} 2002, in prep).

The only component of our Galaxy which could plausibly contain a
significant contribution from stars formed in accreted dwarf galaxies
is the halo, and it contains only about 1\% of the stellar mass of our
Galaxy (e.g., Morrison 1993), and only a fraction of this, the
outer-halo ($\sim$~10\%), could plausibly include stars accreted from
dwarf galaxies (e.g., Unavane, Wyse \& Gilmore 1996).  The outer-halo
stars typically have different properties from the inner-halo which
include relatively lower $\alpha$-abundances, more consistent with
those observed in dwarf galaxies (Nissen \& Schuster 1997).  However,
the kinematic properties of these stars are not always consistent with
what we see in satellite galaxies today (e.g., Gilmore \& Wyse
1998). The samples of well studied halo stars are still too small to
be reliably quantitative or conclusive on any of these issues.

\subsection{Did the ``First Stars'' form in dSph?}

In CDM cosmology dwarf galaxy sized objects are the first structures
to collapse in the early Universe. These first objects, arguably
proto-galaxies, are predicted to have a total mass (dark + baryonic
matter) of $\sim10^6\Msun$ (White \& Rees 1978; Tegmark {\it et al.}
1997).  Within the CDM paradigm most of these small structures will
merge to form much larger galaxies, such as our own, but a fraction may
be supposed to survive until today in something like their original form.
Within these first objects the first generation of stars in the
Universe may form (zero metallicity), perhaps as one massive star
(Abel, Bryan \& Norman 2002; Bromm, Coppi \& Larson 2002).  So far it
has been difficult to constrain the final mass of the ``first''
stars. Current predictions place them between 30 and 300$\Msun$. Such
a large uncertainty makes reliable nucleosynthetic predictions
difficult.

These first collapsed proto-galaxies at 10$^6\Msun$ have a mass which
in most models is very close to the limit where star formation results
in significant blow-out and disruption of the ISM and the total
blow-away of all the gas in the galaxy in a SN explosion ({\it e.g.},
Larson 1974; Dekel \& Silk 1986; Mac Low \& Ferrara 1999; Ferrara \&
Tolstoy 2000; Mori, Ferrara \& Madau 2002).  Thus, the formation (and
subsequent SN explosion) of high mass stars (of order 100$\Msun$) in
such small halos in the early Universe has also been recently invoked
as a convenient way to pollute the early Universe with metals ({\it
e.g.}, Madau, Ferrara \& Rees 2001), and explain the relatively high
abundance of metals in the IGM at quite high redshift, remove the
G-dwarf problem in our Galaxy, and even to reionise the Universe ({\it
e.g.}, Oh {\it et al.} 2001).

As a means of testing the idea that very massive stellar objects
(VMOs) may have exploded at early times in small dwarf galaxy like
objects which have survived until the present day as dSph, we look for
their nucleosynthetic signature in the spectra of the oldest stars our
sample, as predicted by Heger \& Woosley (2002).  We are not looking
directly for Population III objects, but we are looking for any sign
that the stars we observe were polluted by VMOs in the early
Universe. In Figure~\ref{cosmo} we plot the comparison of the
abundances we observe in the only predominantly old galaxy in our
sample, the Sculptor dSph with the predictions of Heger \& Woosley for
the SN products of a zero metallicity VMO.  We also plot as a
comparison the average abundance pattern seen in the most metal poor
halo stars (in the range $-3 <$ [Fe/H] $< -4$, from McWilliam {\it et
al.} 1995).

The Sculptor dSph abundance patterns are very halo-like despite having
significantly higher [Fe/H] and lower average [$\alpha$/Fe] not
obvious in this plot, where all abundances are relative to
$\alpha$-element, Mg.  Comparing the Sculptor abundance patterns with
those predicted by the VMO pair-production SN (progenitor masses
between 140$\Msun$ and 260$\Msun$) models of Heger \& Woosley we can't
find a good match.  It is likely that all the stars in our sample have
been enriched by more than one SN, although perhaps not many
more (see Paper~I).  Most critically we always find elements heavier
than Zn in all our spectra, which are absent in the Heger \& Woosley
VMO pair-production SN.  This does place limits on the models which
anticipate only one zero metallicity SN per proto-galaxy, but a lot
more data are required to be conclusive.

The initial enrichment of the ISM in these galaxies is likely to
develop extremely quickly as a result of the first star(s) formation.
The time scale for SNII, is of order $<10^6$yrs, so the initial
enrichment time scale of the ISM is well beyond the time resolution of
our CMD analysis. Since we are only looking at 2$-$5 stars per galaxy
we are not very likely to find a sample which represents the earliest
most rapid phases of ISM enrichment. We need to observe abundances for
a much larger sample of stars in these dSph to build up an accurate
picture of how the early enrichment may have progressed on the very
short timescales of the initial enrichment.  It is possible, at least
for dSph, that varying rates of loss of gas and metals in SN driven
winds in the earliest starburst phase of formation make our
understanding of the abundance patterns in the oldest stars very
complex.

\newpage
\section{Conclusions}

We have obtained high resolution VLT UVES spectra for 15 RGB stars in
Sculptor, Fornax, Carina and Leo~I dSph galaxies. Adding this to the
previous Keck HIRES observations of 17 RGB stars in Draco, Ursa Minor
and Sextans dSph by SCS01 we have a sizable sample of stars with
which to study the chemical evolution of our smallest neighbors. We
have measured abundances for more than twenty different elements in
all these stars, and used these many different indicators to try and
constrain our understanding of chemical evolution in dwarf
galaxies back to the earliest times.

Our results, similar to SCS01, show a very uniform abundance pattern
across all these galaxies.  This is evidence for very similar chemical
evolution histories in all dSph galaxies, similar initial conditions
and a similar IMF.  It was initially a surprise that the abundance
patterns were so similar since these galaxies have often quite
different star formation histories, and all of the galaxies have
evidence for fairly large spreads in [Fe/H] (see Table~5).  The fact
that these galaxies have always had very small (but variable) rates of
star formation is consistent the patterns we see in the $\alpha$-,
Fe-Peak and Heavy Element abundances.  These variations are fairly
well behaved.  We also find that the global increase in [Fe/H] with
time does not require (or indicate) any significant infall or outflow
of gas or metals from these systems, but this is very dependent on the
accuracy of the models used to determine yields.

A small star formation rate could also have the effect of limiting the
impact of massive SNII on the chemical evolution of these galaxies,
effectively truncating the IMF, which could account for the marked
difference between the abundance patterns seen in dSph stars and those
found in our Galaxy. This data set allows us to infer more clearly the
effects of AGB star mass loss and low mass SNII on the chemical
evolution of a galaxy.  In our Galaxy detailed signatures of this kind
of enrichment may often be ``washed out'' over time, possibly because
the chemical properties of a system which has high mass SNII will (not
surprisingly) be totally dominated by them.  It might also be possible
to explain the abundance patterns in terms of mass (and metal) loss in
dwarf galaxies, but for this (admittedly small) data set this appears
to be a contrived solution as, from current models, this would have to
preferential mass loss of massive star ($>20\Msun$) SNII products.

The observations we have presented here also add to the evidence which
suggests that the abundance patterns observed in stars in small nearby
galaxies are markedly different to those observed in our Galaxy. This
makes it difficult for a significant proportion of the stars observed
in our Galaxy to have been formed in small galaxies which subsequently
merged to form the Milky Way. This is true of the disk, the bulge and
inner-halo of our Galaxy from the (limited) data available to date.
The only component of our Galaxy with some similarity to the stellar
populations found in dwarf galaxies is the outer-halo, which contains
a very small fraction by mass of stars in our Galaxy.

With our accurate measurements of a range of different elements we
have found that we can plot the enrichment of the ISM in these
galaxies due to the observed star formation rate variations 
directly. This is especially evident in the alpha elements (see
Figure~\ref{alf5}). This is the first time such a direct {\it
measurement} has been made of the chemical evolution of a galaxy over
nearly a Hubble time. Currently the number of stars observed in each
dSph is too small to be able to make concrete statements, but it is
clear that larger samples coming from instruments like FLAMES, are
going to have a dramatic impact upon our understanding of the chemical
evolution of galaxies.  These larger samples will also allow us to be
more definite in our interpretation of the nucleosynthetic origin of
the elements used to trace SNII and SNIa masses and rates, as well as
the timescales and effect of AGB stars.

There are many good arguments to suggest that star formation has
occurred in these small objects back to the earliest times in the
Universe. There are also suggestions that the first stars were formed
in dwarf galaxy sized objects. Nonetheless we can find no evidence
that any of the stars we observed were formed from an ISM enriched by
the single zero metallicity stars proposed in the literature.

The absolute ages of the oldest stars can still not be determined very
accurately, a reasonable error of $\sim$2~Gyr is the time elapsed
between z=6 and z=2. But with abundances we can measure the enrichment
of the gas in a galaxy which provides enough information to allow us
to determine a more accurate absolute time scale.  The galaxy must
produce SN at a rate consistent with the star formation history.  We
have taken this approach here for a limited sample of stars with
promising results, we look forward to significantly extending our
sample in the not too distant future.

\acknowledgements{
{\bf Acknowledgments:} We thank the Paranal Observatory staff for the
excellent support we received during both of our visitor runs.  We
also acknowledge useful discussions with Tom Abel, Andrea Ferrara,
Alex Heger, Sally Oey, Evan Skillman, Rosie Wyse and Sukyoung Yi.  
We also thank the anonymous referee for helpful comments. ET
gratefully acknowledges support from a fellowship of the Royal
Netherlands Academy of Arts and Sciences, and PATT travel support from
the University of Oxford. KAV would like to thank the NSF for support
through a CAREER award, AST-9984073.}

\newpage
\def\x{\enspace}
\def\xx{\enspace\enspace}
\def\xxx{\enspace\enspace\enspace}
\def\xxxx{\enspace\enspace\enspace\enspace}
\def\xxxxx{\enspace\enspace\enspace\enspace\enspace}
\def\jref#1 #2 #3 #4 {{\par\noindent \hangindent=3em \hangafter=1
      \advance \rightskip by 5em #1, {\it#2}, {\bf#3}, {#4} \par}}
\def\ref#1{{\par\noindent \hangindent=3em \hangafter=1
      \advance \rightskip by 5em #1 \par}}
\def\endtable{\endgroup}
\def\tableheight{\vrule width 0pt height 8.5pt depth 3.5pt}
{\catcode`|=\active \catcode`&=\active
    \gdef\tabledelim{\catcode`|=\active \let|=\vbar
                     \catcode`&=\active \let&=\nobar} }
\def\table{\begingroup
    \def\twidth{\hsize}
    \def\tablewidth##1{\def\twidth{##1}}
    \def\defaultheight{\vrule width 0pt height 8.5pt depth 3.5pt}
    \def\heightdepth##1{\dimen0=##1
        \ifdim\dimen0>5pt
            \divide\dimen0 by 2 \advance\dimen0 by 2.5pt
            \dimen1=\dimen0 \advance\dimen1 by -5pt
            \vrule width 0pt height \the\dimen0  depth \the\dimen1
        \else  \divide\dimen0 by 2
            \vrule width 0pt height \the\dimen0  depth \the\dimen0 \fi}
    \def\spacing##1{\def\defaultheight{\heightdepth{##1}}}
    \def\nextheight##1{\noalign{\gdef\tableheight{\heightdepth{##1}}}}
    \def\end{\cr\noalign{\gdef\tableheight{\defaultheight}}}
    \def\zerowidth##1{\omit\hidewidth ##1 \hidewidth}
    \def\hline{\noalign{\hrule}}
    \def\skip##1{\noalign{\vskip##1}}
    \def\bskip##1{\noalign{\hbox to \twidth{\vrule height##1 depth 0pt \hfil
        \vrule height##1 depth 0pt}}}
    \def\header##1{\noalign{\hbox to \twidth{\hfil ##1 \unskip\hfil}}}
    \def\bheader##1{\noalign{\hbox to \twidth{\vrule\hfil ##1
        \unskip\hfil\vrule}}}
    \def\spanloop{\span\omit \advance\mscount by -1}
    \def\extend##1##2{\omit
        \mscount=##1 \multiply\mscount by 2 \advance\mscount by -1
        \loop\ifnum\mscount>1 \spanloop\repeat \ \hfil ##2 \unskip\hfil}
    \def\vbar{&\vrule&}
    \def\nobar{&&}
    \def\hdash##1{ \noalign{ \relax \gdef\tableheight{\heightdepth{0pt}}
        \toks0={} \count0=1 \count1=0 \putout##1\end
        \toks0=\expandafter{\the\toks0 &\end} \xdef\piggy{\the\toks0} }
        \piggy}
    \let\e=\expandafter
    \def\putspace{\ifnum\count0>1 \advance\count0 by -1
        \toks0=\e\e\e{\the\e\toks0\e&\e\multispan\e{\the\count0}\hfill}
        \fi \count0=0 }
     \def\putrule{\ifnum\count1>0 \advance\count1 by 1
        \toks0=\e\e\e{\the\e\toks0\e&\e\multispan\e{\the\count1}\leaders\hrule\
hfill}
        \fi \count1=0 }
    \def\putout##1{\ifx##1\end \putspace \putrule \let\next=\relax
        \else \let\next=\putout
            \ifx##1- \advance\count1 by 2 \putspace
            \else    \advance\count0 by 2 \putrule \fi \fi \next}   }
\def\tablespec#1{
    \def\vdimens{\noexpand\tableheight}
    \def\tabby{\tabskip=0pt plus100pt minus100pt}
    \def\r{&################\tabby&\hfil################\unskip}
    \def\c{&################\tabby&\hfil################\unskip\hfil}
    \def\l{&################\tabby&################\unskip\hfil}
    \edef\templ{\noexpand\vdimens ########\unskip  #1
         \unskip&########\tabskip=0pt&########\cr}
    \tabledelim
    \edef\body##1{ \vbox{
        \tabskip=0pt \offinterlineskip
        \halign to \twidth {\templ ##1}}} }

\centerline{\bf Table 1: The Dwarf Spheroidal Galaxies}
\vskip-1cm

$$
\footnotesize
\table
\tablespec{\l\l\l\c\c\c\c\l\l}
\body{
\skip{0.06cm}
\hline
\skip{0.025cm}
\hline
\skip{.2cm}
& Object & l & b & (m-M)$_0$ & E(B$-$V) & M$_V$ & $\Sigma_0$    & v$_r$ &M$_{tot}$      &\end
&        &   &   &     mag   &   mag    &  mag  & mag/arcsec$^2$&km/s   &10$^6\Msun$  &\end
\skip{.1cm}
\hline
\skip{.05cm}
\hline
\skip{.45cm}
& Sculptor&287.5 &$-$83.2 &19.54$\pm$0.08&0.02$\pm$0.02&$-$11.1&23.7$\pm$0.4&110$\pm$3&6.4 &\end      
& Carina  &260.1 &$-$22.2 &20.03$\pm$0.09&0.025$\pm$0.02&$-$9.3 &25.5$\pm$0.4&224$\pm$3&13  &\end
& Leo~I   &226.0 &$+$49.1 &21.99$\pm$0.20&0.01$\pm$0.01&$-$11.9&22.4$\pm$0.3&286$\pm$2&22  &\end
& Fornax  &237.1 &$-$65.7 &20.70$\pm$0.12&0.03$\pm$0.01&$-$13.2&23.4$\pm$0.3& 53$\pm$3&68  &\end
\skip{.2 cm}
& Draco   & 86.4 &$+$34.7 &19.58$\pm$0.15&0.03$\pm$0.01&$-$8.8 &25.3$\pm$0.5&$-$293$\pm$2&22  &\end
& Ursa Minor&105.0&$+$44.8&19.11$\pm$0.10&0.03$\pm$0.02&$-$8.9 &25.5$\pm$0.5&$-$248$\pm$2&23  &\end
& Sextans &243.5 & $+$42.3&19.67$\pm$0.08&0.03$\pm$0.01&$-$9.5 &26.2$\pm$0.5&227$\pm$3&19  &\end
\skip{.25cm}
\hline
\skip{0.025cm}
\hline
}
\endtable
$$
\noindent{all these data are taken from Mateo 1998}

\newpage
\centerline{\bf Table 2: Isochrone Ages \& $\alpha$-Abundances in UVES Dwarf Spheroidal stars}
\vskip-1cm

$$
\scriptsize
\table
\tablespec{\l\l\l\l\c\c\c\c}
\body{
\skip{0.06cm}
\hline
\skip{0.025cm}
\hline
\skip{.2cm}
& Galaxy & star id &age   &[FeI/H] & [O/Fe] & [Mg/Fe] & [Ca/Fe] & [$\alpha$/Fe] &\end
&        &         &(Gyr) &        &        &         &         &               &\end
\skip{.1cm}
\hline
\skip{.05cm}
\hline
\skip{.45cm}
& Sculptor & H459  & 15$^{+1}_{-3}$ &$-1.66\pm0.02$&$0.12\pm0.15$ &$0.36\pm0.13$ &$0.24\pm0.05$ &0.18    &\end
&          & H479 &  15$^{+1}_{-0}$ &$-1.77\pm0.02$&$0.38\pm0.11$ &$0.26\pm0.16$ &$0.17\pm0.05$ &0.13    &\end
&          & H461 &  15$^{+1}_{-0}$ &$-1.56\pm0.02$&$0.34\pm0.16$ &$0.18\pm0.11$ &$0.22\pm0.06$ &0.13    &\end
&          & H482 &  10$^{+5}_{-2}$ &$-1.24\pm0.02$&$0.08\pm0.18$ &$0.09\pm0.13$ &$0.06\pm0.06$ &$-$0.01 &\end
&          & H400 &  15$^{+1}_{-0}$ &$-1.98\pm0.03$& --           &$0.37\pm0.12$ &$0.38\pm0.09$ &0.23    &\end
& Fornax   & M12  &  15$^{+1}_{-3}$ &$-1.60\pm0.02$&$0.12\pm0.18$ &$0.09\pm0.07$ &$0.23\pm0.06$ & 0.12   &\end
&          & M25  &  10$^{+5}_{-3}$ &$-1.21\pm0.02$&$0.07\pm0.11$ &$0.02\pm0.09$ &$0.21\pm0.06$ & 0.03   &\end
&          & M21  &  2$^{+0.1}_{-1}$&$-0.67\pm0.03$&$0.02\pm0.17$ &$0.20\pm0.12$ &$0.23\pm0.08$ & 0.27   &\end
& Leo~I    & M5   &  8$^{+3}_{-3}$  &$-1.52\pm0.02$&$0.43\pm0.13$ &$0.12\pm0.12$ &$0.15\pm0.06$ & 0.13   &\end
&          & M2 &  2$^{+0.5}_{-1}$  &$-1.06\pm0.02$&$-0.04\pm0.13$&$-0.19\pm0.19$&$0.02\pm0.06$ &$-$0.08 &\end
& Carina   & M4   &  11$^{+4}_{-3}$ &$-1.59\pm0.02$&$0.12\pm0.09$ &$ 0.26\pm0.09$&$ 0.14\pm0.04$& 0.14   &\end
&          & M3   &  13$^{+3}_{-3}$ &$-1.65\pm0.02$&$-0.06\pm0.12$&$-0.27\pm0.12$&$-0.10\pm0.06$&$-$0.26 &\end
&          & M2   &  10$^{+2}_{-2}$ &$-1.60\pm0.02$&$0.34\pm0.12$ &$ 0.23\pm0.10$&$ 0.20\pm0.05$& 0.17   &\end
&          & M12  &  5$^{+2}_{-1}$  &$-1.41\pm0.02$&$0.07\pm0.10$ &$ 0.24\pm0.10$&$ 0.12\pm0.05$& 0.13   &\end
&          & M10  &  15$^{+1}_{-0}$ &$-1.94\pm0.02$&$-0.02\pm0.19$&$ 0.06\pm0.11$&$-0.02\pm0.05$& 0.05   &\end
\skip{.25cm}
\hline
\skip{0.025cm}
\hline
}
\endtable
$$

\newpage

\centerline{\bf Table 3: Isochrone Ages for Dwarf Spheroidal stars measured by SCS01}
\vskip-1cm
$$
\table
\tablespec{\l\r\r\c\r}
\body{
\skip{0.06cm}
\hline
\skip{0.025cm}
\hline
\skip{.2cm}
& Galaxy & star id &age (Gyr) &[FeI/H] & [$\alpha$/Fe] &\end
\skip{.1cm}
\hline
\skip{.05cm}
\hline
\skip{.45cm}
& Draco    & 11    & 9  & $-1.72\pm0.11$ & $0.11\pm0.05$  &\end
&	   & 343   & 10 & $-1.86\pm0.11$ & $0.01\pm0.07$  &\end
&          & 473   & 6  & $-1.44\pm0.07$ & $0.09\pm0.07$  &\end
&          & 267   & 15 & $-1.67\pm0.13$ & $0.15\pm0.07$  &\end
&          & 24    & 15 & $-2.36\pm0.09$ & $0.08\pm0.05$  &\end 
&          & 119   & 15 & $-2.97\pm0.15$ & $0.11\pm0.06$  &\end 
& Ursa Minor & 177  & 15& $-2.01\pm0.11$ & $0.19\pm0.05$  &\end
&            & 297  & 15& $-1.68\pm0.11$ & $0.10\pm0.05$  &\end 
&            & K(C2)& 15& $-2.17\pm0.12$ & $0.25\pm0.07$  &\end
&            & O    & 15& $-1.91\pm0.11$ & $0.17\pm0.05$  &\end
&            & 199  & 15& $-1.40\pm0.11$ &$-0.03\pm0.06$  &\end
&            & 168  & 15& $-2.18\pm0.12$ & $0.10\pm0.10$  &\end
& Sextans    & 35   & 15& $-1.93\pm0.11$ & $0.13\pm0.06$  &\end
&            & DC5  & 15& $-1.93\pm0.11$ & $0.12\pm0.07$  &\end
&            & 49   & 15& $-2.85\pm0.13$ &$-0.01\pm0.11$  &\end 
&            & 58   & 6 & $-1.45\pm0.12$ &$-0.26\pm0.07$  &\end
&            & 36   & 15& $-2.19\pm0.12$ & $0.10\pm0.07$  &\end
\skip{.25cm}
\hline
\skip{0.025cm}
\hline
}
\endtable
$$

\newpage

\centerline{\bf Table 4: Heavy Element and Iron-Peak Abundances in UVES Dwarf Spheroidal stars}
\vskip-1cm

$$
\table
\tablespec{\l\r\r\c\c\c}
\body{
\skip{0.06cm}
\hline
\skip{0.025cm}
\hline
\skip{.2cm}
& Galaxy & star id & [Ba/Eu] & [Mn/Fe] & [Zn/Fe] &[Cu/$\alpha$]&\end
&        &         &         &         &         &             &\end
\skip{.1cm}
\hline
\skip{.05cm}
\hline
\skip{.45cm}
& Sculptor & H459  &$-0.30$&$-0.52$&$ 0.27$  &$-1.23$ &\end
&          & H479  &$-0.44$&$-0.52$&$-0.38$ &$-0.96$ &\end
&          & H461  &$-0.14$&$-0.62$&$-0.33$ &$-0.92$ &\end
&          & H482  &0.03   &$-0.39$&$ 0.08$ &$-1.12$ &\end
&          & H400  &$-0.27$&--     & --   &$<-0.69$&\end
& Fornax   & M12   &$-0.31$&$-0.47$&$-0.29$ &$-0.79$ &\end
&          & M25   &  0.23 &$-0.43$&$ 0.00$ &$-0.63$ &\end
&          & M21   &  0.32 &$-0.61$& --   &  0.12  &\end
& Leo~I    & M5    &$-0.39$&$-0.48$&$-0.31$ &$-0.73$ &\end
&          & M2    &$-0.25$&$-0.31$&$-1.46$ &$-0.47$ &\end
& Carina   & M4    &$-0.17$&$-0.46$&$-0.10$ &$-0.77$ &\end
&          & M3    &$-0.19$&$-0.18$&$-0.30$ &$-0.59$ &\end
&          & M2    & 0.04  &$-0.50$&$0.04$  &$-0.80$ &\end
&          & M12   & 0.07  &$-0.45$&$-0.22$ &$-0.74$ &\end
&          & M10   &$-0.55$&$-0.47$&$-0.20$ &$<-0.65$&\end
\skip{.25cm}
\hline
\skip{0.025cm}
\hline
}
\endtable
$$

\newpage

\centerline{\bf Table 5: Metallicity Spreads: photometric versus spectroscopic}
\vskip-1cm

$$
\table
\tablespec{\l\l\l\l\l\l}
\body{
\skip{0.06cm}
\hline
\skip{0.025cm}
\hline
\skip{.2cm}
& Galaxy & Photometric & & Spectroscopic & &\end
&        & $<[Fe/H]>$ & $\delta[Fe/H]$ & $<[Fe/H]>$ & $\delta[Fe/H]$ &\end
\skip{.1cm}
\hline
\skip{.05cm}
\hline
\skip{.45cm}
& Ursa Minor& -2.2  & 0.1  & -1.9 & 1.4 & SCS01\end
& Draco     & -2.1  & 0.2  & -2.0 & 1.5 & SCS01\end
& Carina    & -2.0  & 0.2  & -1.6 & 0.5 & this work \end
& Sculptor  & -1.8  & 0.1  & -1.5 & 0.9 & this work \end
& Sextans   & -1.7  & 0.2  & -2.1 & 0.7 & SCS01\end
& Leo~I     & -1.5  & 0.4  & -1.3:& 0.5:& this work \end
& Fornax    & -1.3  & 0.2  & -1.2 & 1.0 & this work \end
\skip{.25cm}
\hline
\skip{0.025cm}
\hline
}
\endtable
$$

\newpage

\begin{figure}
\centerline{\hbox{\psfig{figure=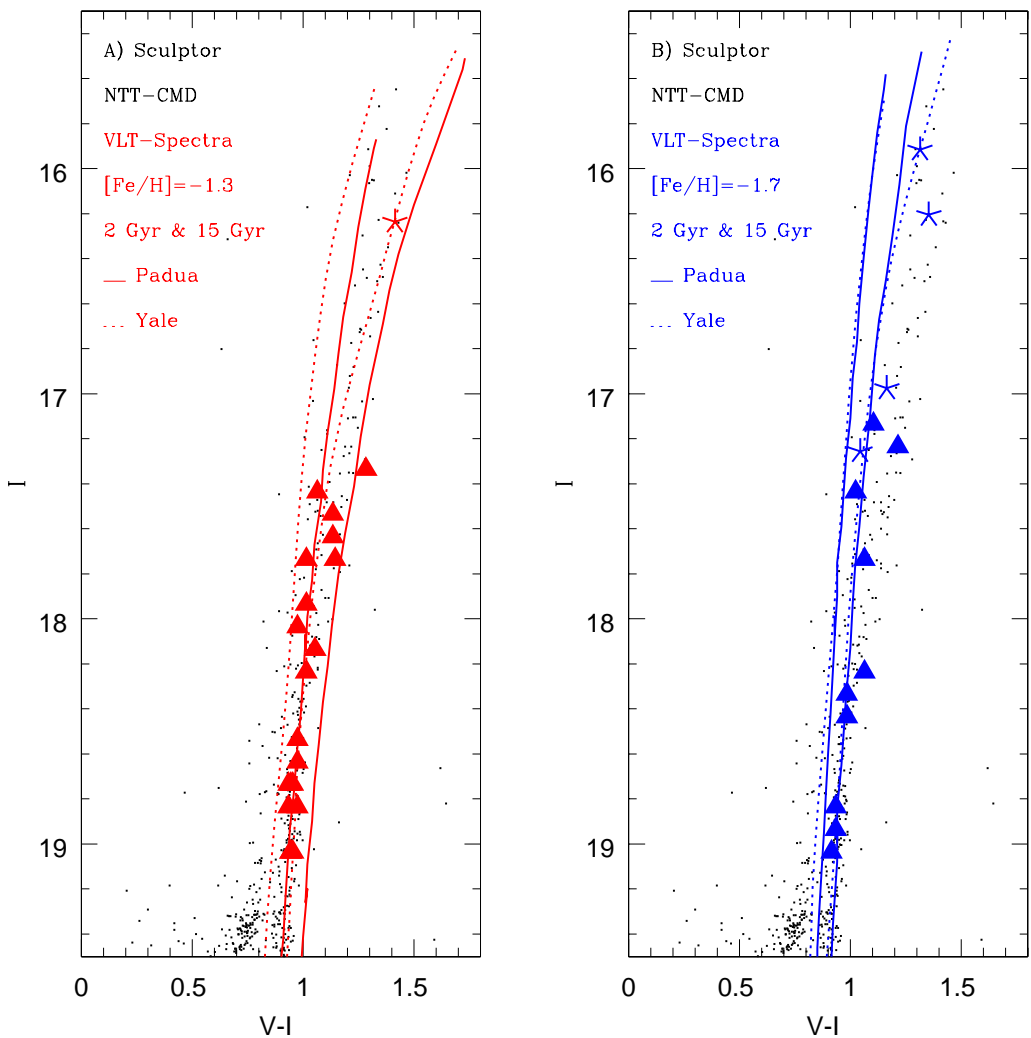,width=15cm}}}
\caption{
The positions of our UVES targets ({\it stars}) and the CaT targets
from T01 ({\it filled triangles}) on the RGB in Sculptor.  Photometry
from NTT images by Tolstoy {\it et al.}, in prep.  Left panel shows
targets with [Fe/H] = $-1.3 \pm 0.3$ and isochrones from Bertelli {\it
et al.} (1994 = Padua) and Yi {\it et al.} (2001 = Yale).  Similarly,
the right panel shows the locations of targets with [Fe/H] = $-1.7 \pm
0.3$.  This figure highlights the difficulties in determining accurate
ages from isochrones even when the metallicity is known.
}
\label{sclprob}
\end{figure}								     

\begin{figure}
\centerline{\hbox{\psfig{figure=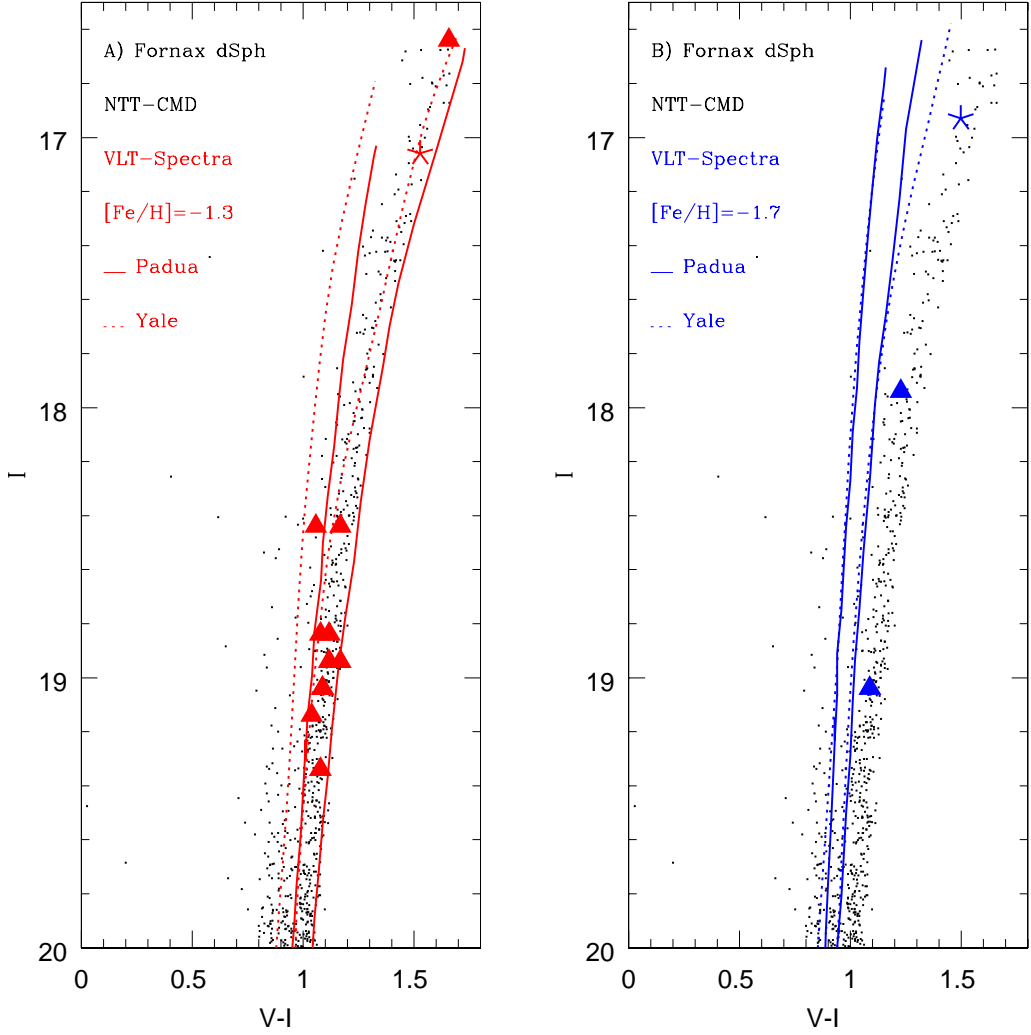,width=15cm}}}
\caption{
Same as Figure 1, but for imaging of and targets in Fornax.
}
\label{fnxprob}
\end{figure}								     

\begin{figure}
\centerline{\hbox{\psfig{figure=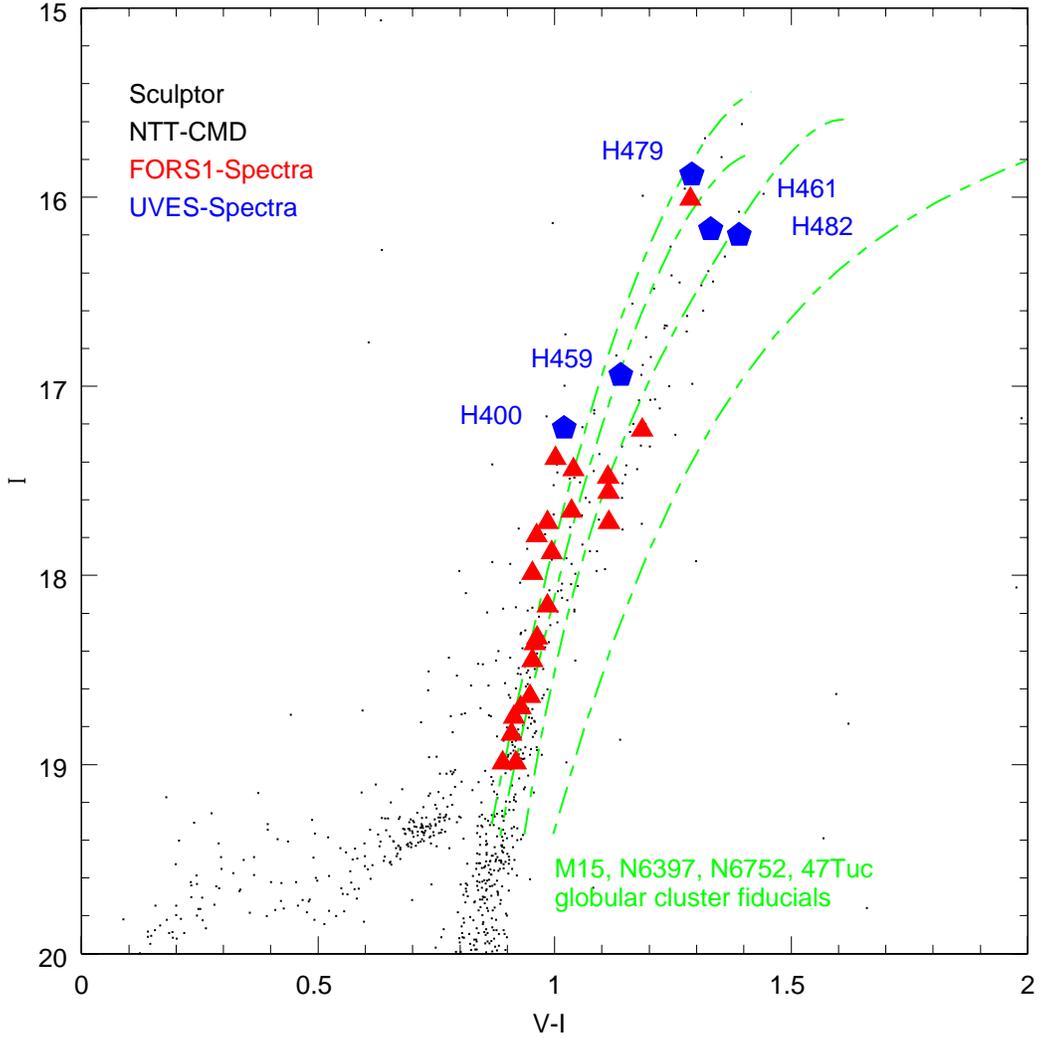,width=15cm}}}
\caption{
Combined CMD for Sculptor from three NTT fields.
All of the CaT ({\it filled triangles}, from T01) and UVES ({\it filled 
pentagons}) targets are marked.   {\it Dashed lines} are the RGB
fiducials for the globular clusters: M15, NGC 6397, NGC 6752, and 
47 Tucanae with metallicities [Fe/H] = $-2.2, -1.9, -1.5, -0.7$, 
respectively, from left to right (from Da Costa \& Armandroff 1990). 
}
\label{sclcmd}
\end{figure}								     
								     
\begin{figure}
\centerline{\hbox{\psfig{figure=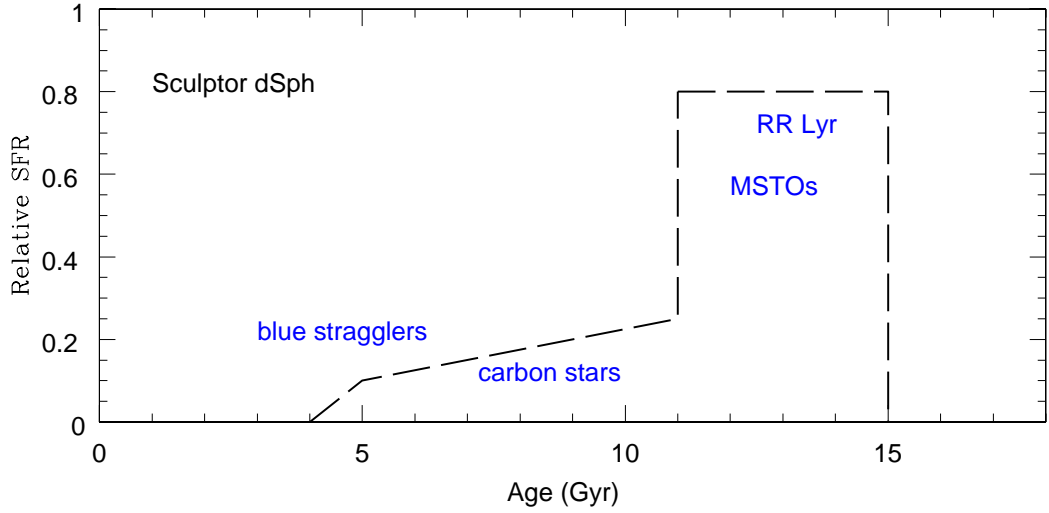,width=15cm}}}
\caption{
A star formation schematic for Sculptor that is consistent with its
stellar population and considering how its star formation rate is
likely to have varied since the epoch of globular cluster formation
around 15 Gyr ago (data sets are discussed in Section 3.3.1).  The
stellar evolution phases used to determine the star formation rate 
at various ages are
labelled.
}
\label{sclsfh}
\end{figure}								     

\begin{figure}
\centerline{\hbox{\psfig{figure=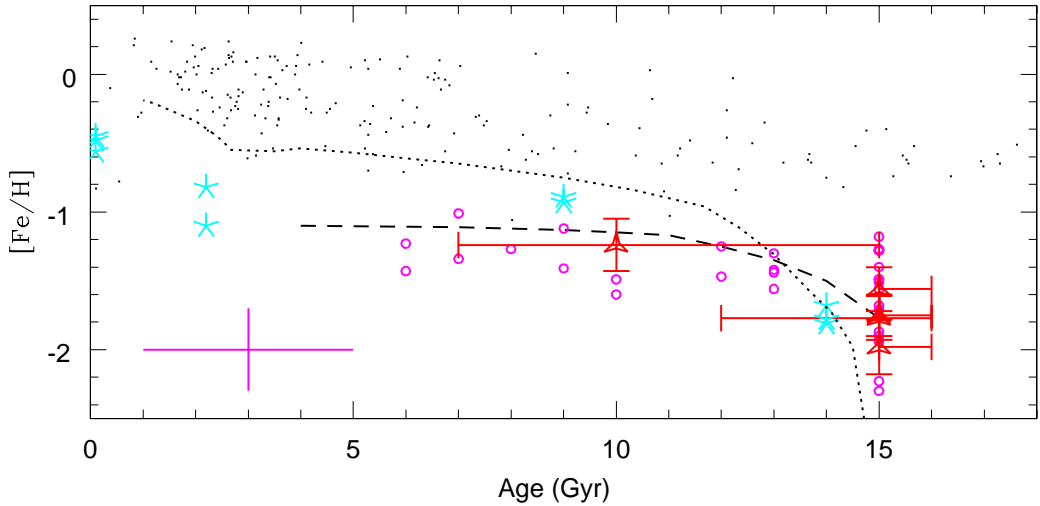,width=15cm}}}
\caption{
A chemical evolution scenario for Sculptor over its entire
star-forming history ($\sim$15 Gyr).  Ages have been determined by
fitting theoretical isochrones to the stars RGB positions, given each
stellar metallicity (discussed in Section 3.2; also see Figure 1).
The dashed line adopts the SFH shown in Figure 4.  The dotted line is
the LMC age-metallicity relation for the bursting model by Pagel \&
Tautvai\v{s}ien\.e (1998).  CaT measurements of metallicity from T01
({\it small circles}) assume [Ca/Fe] = 0; the typical error bar per CaT
observation is noted in the lower left.  Direct iron abundances from
our UVES spectra ({\it filled triangles with error bars}) are in good
agreement with the CaT measurements.  LMC star cluster iron abundances
by H00 ({\it stars}) are offset from the LMC bursting model, but have
similar iron abundances as the Sculptor stars at the oldest ages.  The
numerous small dots show the iron abundances in Galactic disk stars by
E93.
}
\label{sclzfh}
\end{figure}								     

\begin{figure}
\centerline{\hbox{\psfig{figure=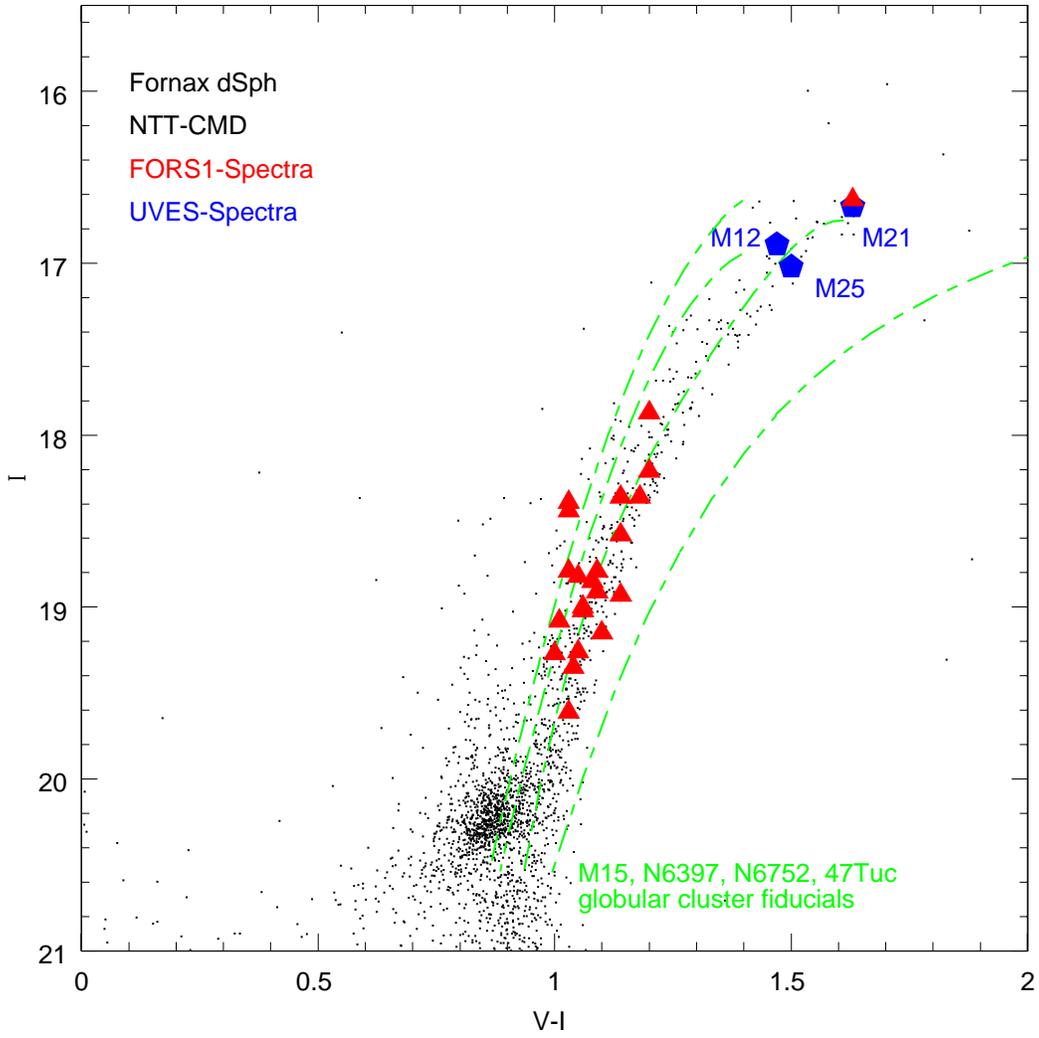,width=15cm}}}
\caption{
Same as Figure 3, but for targets in Fornax.
}
\label{fnxcmd}
\end{figure}								     
								     
\newpage
\clearpage

\begin{figure}
\centerline{\hbox{\psfig{figure=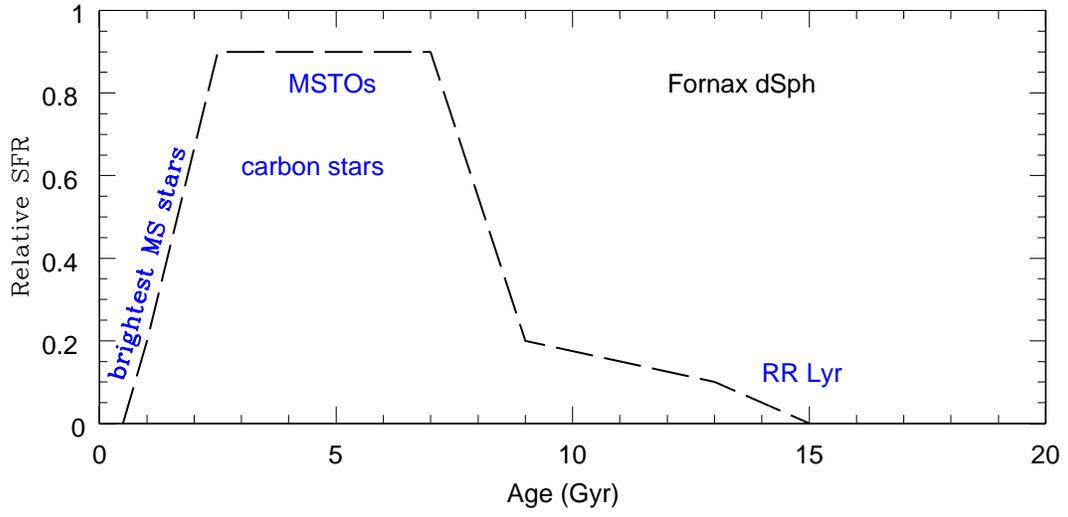,width=15cm}}}
\caption{
Same as Figure 4, but a schematic star formation history for Fornax
(data sets are discussed in Section 3.4.1).
}
\label{fnxsfh}
\end{figure}								     

\begin{figure}
\centerline{\hbox{\psfig{figure=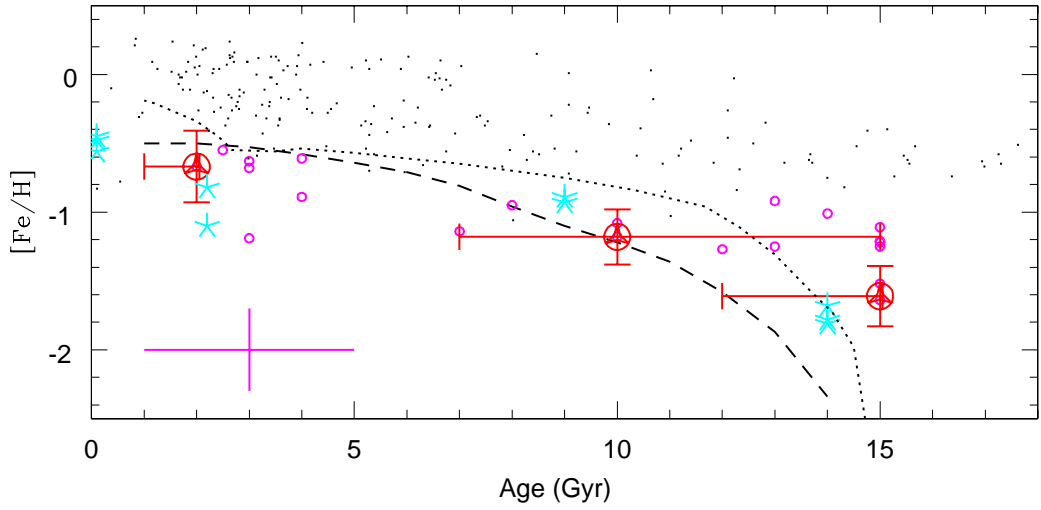,width=15cm}}}
\caption{
Same as Figure 6, but a chemical evolution scenario for Fornax.  The
dashed line adopts the SFH shown in Figure 7.  The iron abundances
from the UVES Fornax targets are marked by {\it filled triangles with
circles} and error bars.
}
\label{fnxzfh}
\end{figure}								     

\begin{figure}
\centerline{\hbox{\psfig{figure=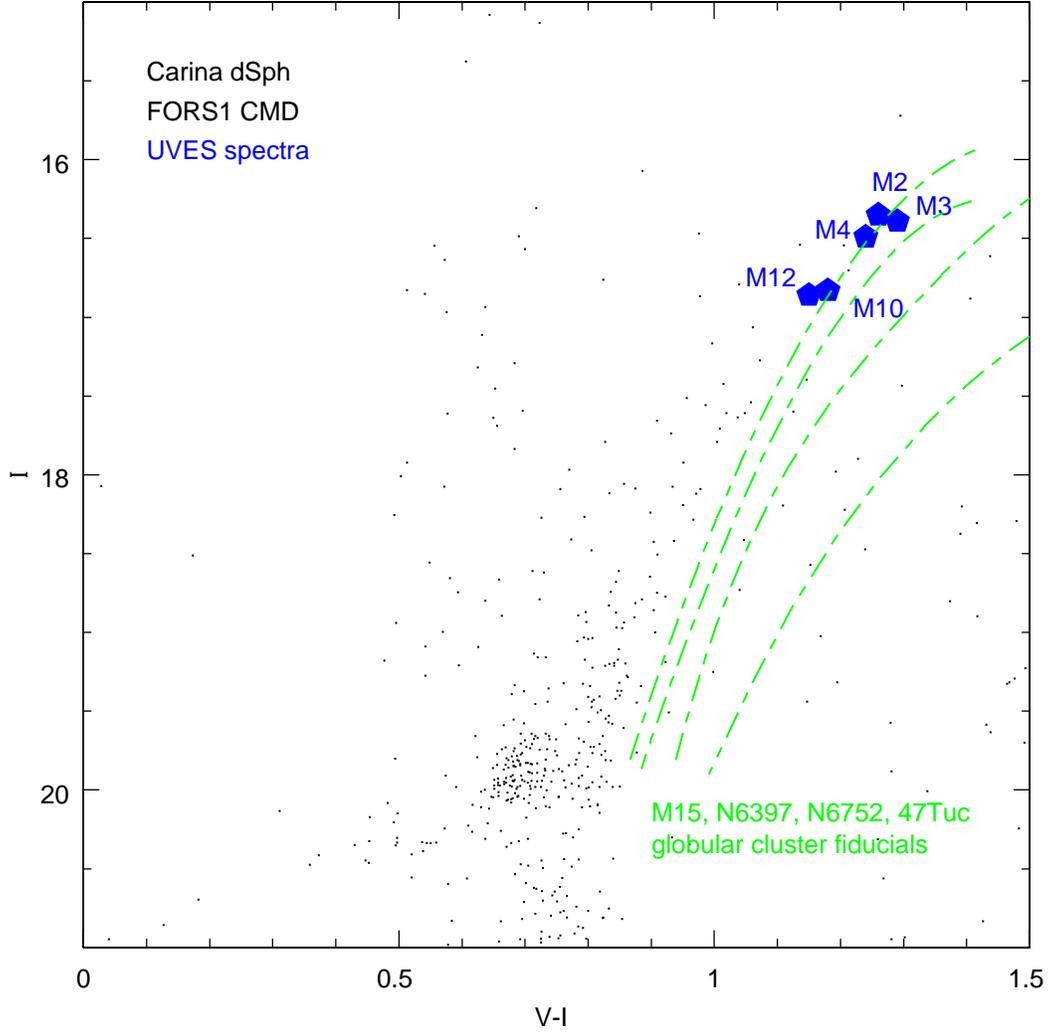,width=15cm}}}
\caption{
Same as Figure 3, but for targets in Carina. The CMD comes from
VLT/FORS1 archival imaging data.
}
\label{carcmd}
\end{figure}								     
								     
\begin{figure}
\centerline{\hbox{\psfig{figure=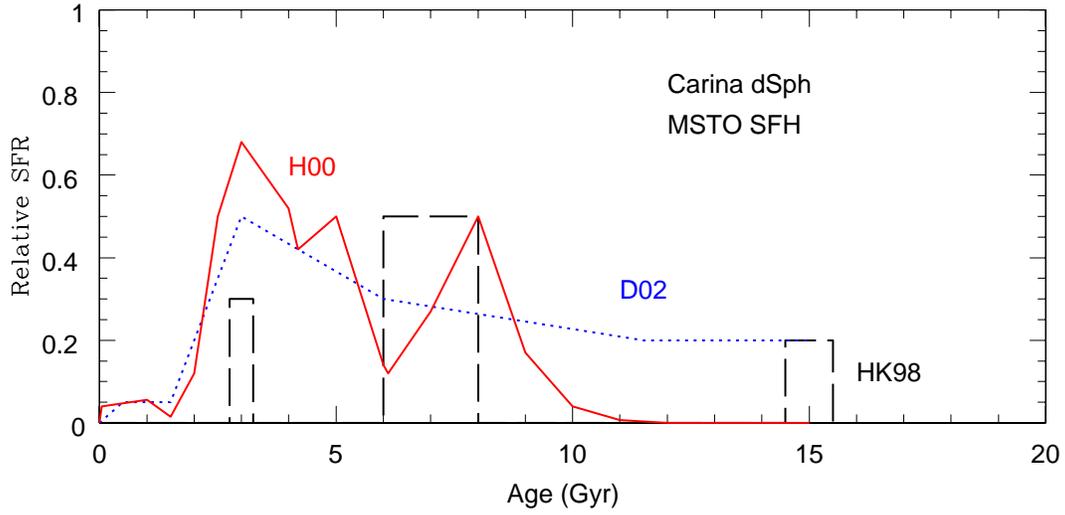,width=15cm}}}
\caption{
Same as Figure 4, but a schematic star formation history for Carina.
Dashed lines from Hurley-Keller {\it et al.} (1998 = HK98) from
wide-field ground-based CTIO imaging.  The solid line (Hernandez {\it
et al.} 2000 = H00) and dotted line (Dolphin 2002 = D02) are based on
models that use the same HST/WFPC2 dataset.  Details on these SFHs are
discussed in Section 3.5.1.
}
\label{carsfh}
\end{figure}								     

\begin{figure}
\centerline{\hbox{\psfig{figure=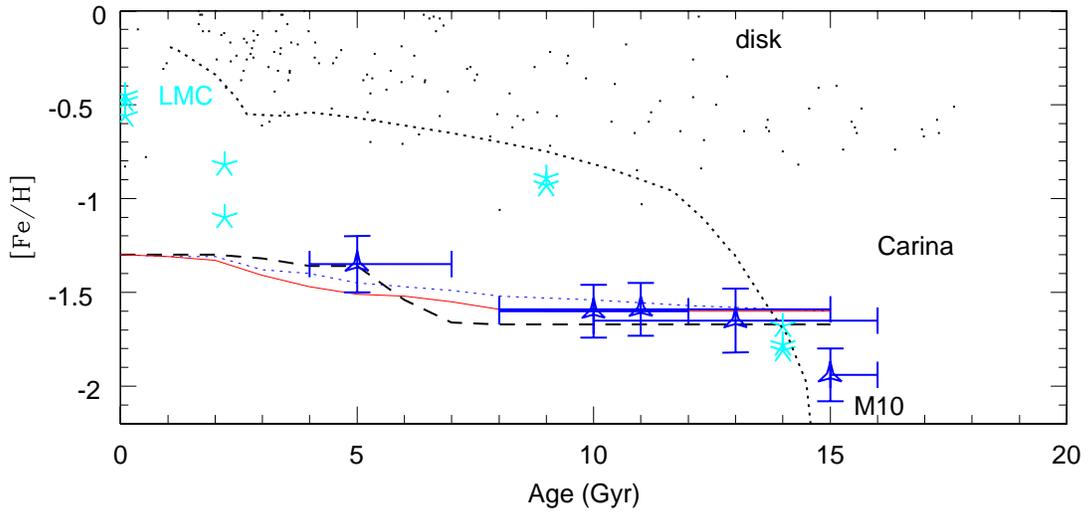,width=15cm}}}
\caption{
Same as Figure 5, but a chemical evolution scenario for Carina.
The grouped solid, dashed, and dotted lines adopt the corresponding
SFHs displayed in Figure 13.   The iron abundances from the UVES
Carina targets are marked by {\it filled triangles} with error bars.
}
\label{carzfh}
\end{figure}								     

\begin{figure}
\centerline{\hbox{\psfig{figure=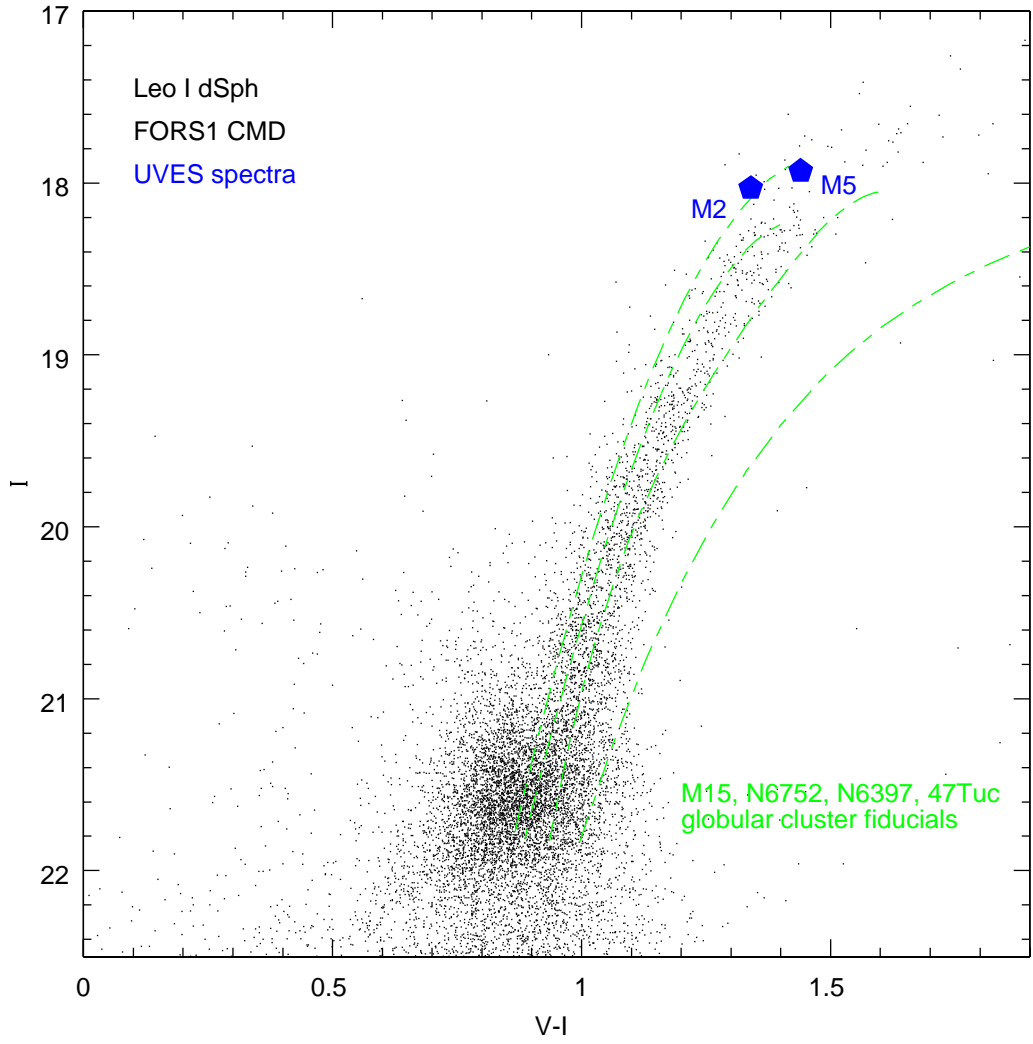,width=15cm}}}
\caption{
Same as Figure 9, but for targets in Leo I.  The CMD comes from
VLT/FORS1 archival imaging data.
}
\label{leocmd}
\end{figure}								     
								     
\begin{figure}
\centerline{\hbox{\psfig{figure=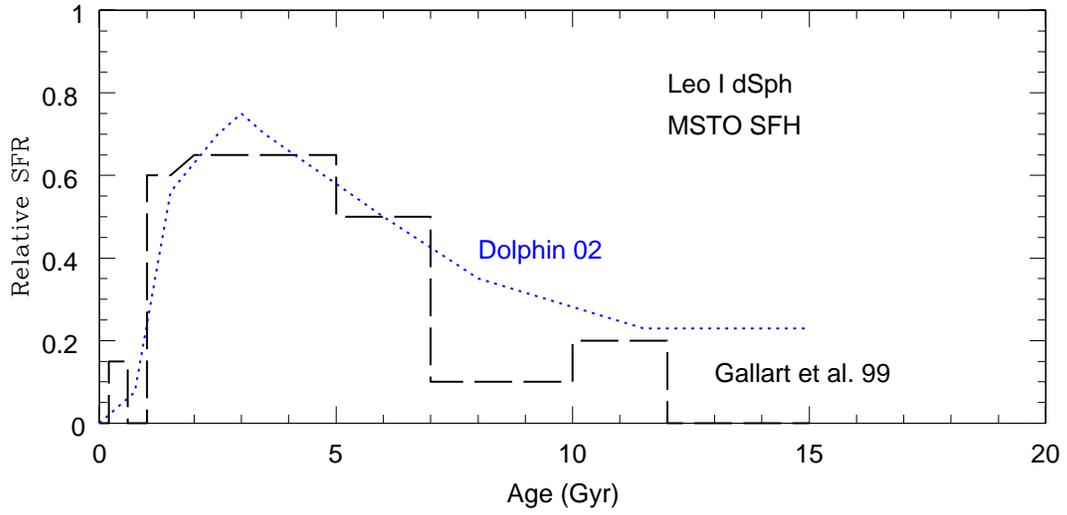,width=15cm}}}
\caption{
Same as Figure 4, but a schematic star formation history for Leo I 
Dashed line (Gallart {\it et al.} 1999) and dotted line (Dolphin 2002) are
based on the same HST/WFPC2 data.   Details on these SFHs are
discussed in Section 3.6.1.
}
\label{leosfh}
\end{figure}								     

\begin{figure}
\centerline{\hbox{\psfig{figure=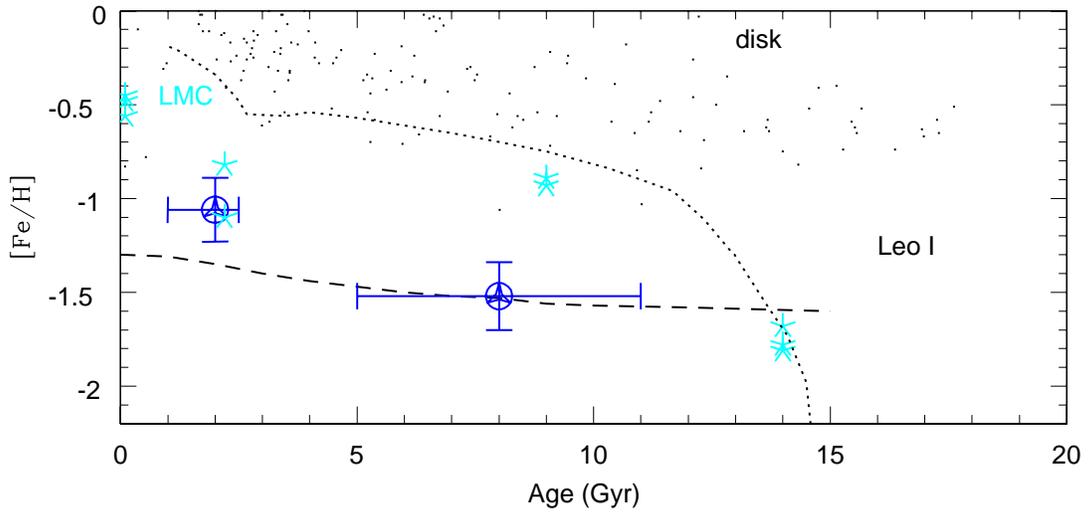,width=15cm}}}
\caption{
Same as Figure 6, but a chemical evolution scenario for Leo I.  The
SFHs in Figure 13 are so similar that they produce almost the same
chemical evolution history ({\it dashed line}).  The iron abundances
from the UVES Leo I targets are marked by {\it filled triangles with
circles} with error bars.
}
\label{leozfh}
\end{figure}								     

\newpage
\clearpage

\begin{figure}
\centerline{\hbox{\psfig{figure=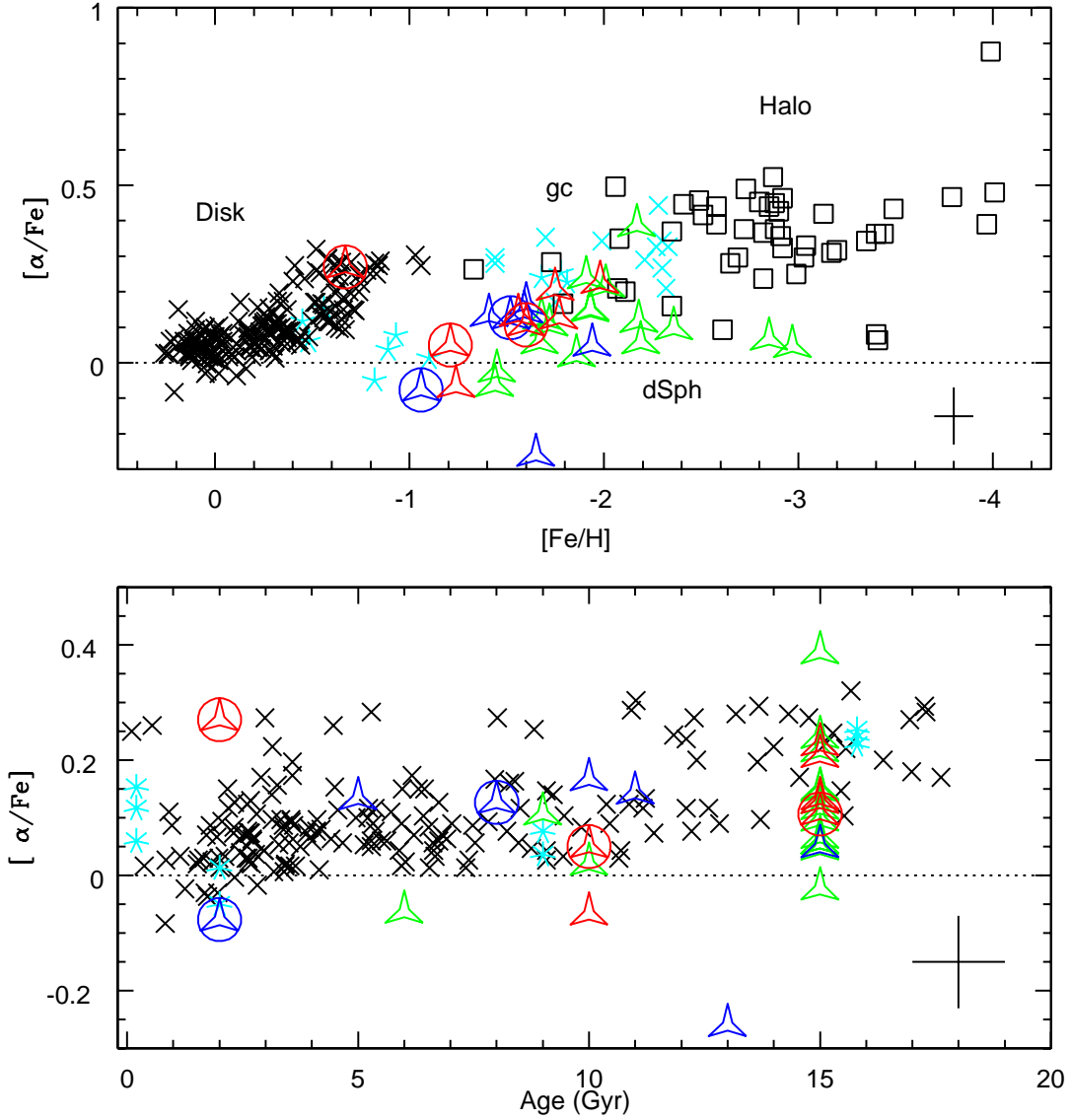,width=15cm}}}
\caption{
$\alpha$-Element abundances, where [$\alpha$/Fe] = 1/3 ([Mg/Fe] +
[Ca/Fe] + [Ti/Fe]).  Our UVES $\alpha$-abundances are plotted versus
[Fe/H] in the upper panel and versus age determined from isochrone
analysis in the lower panel.  The symbols are: blue triangles are the
Carina data; blue triangle plus circle are Leo~I data; red triangles
are Sculptor data; and the red triangle plus circle are Fornax data.
The green triangles are data on Draco, Ursa Minor, Sextans from SCS01.
The black crosses are Galactic disk star measurements from E93; the
open squares are halo data from McWilliam et al. 1995 and the light
blue stars are UVES data from a study of LMC star clusters of
different ages from H00. There are also light blue crosses which are
our Galactic globular cluster measurements, combined with those of
SCS01.  We have plotted an average representative set of error bars in
each plot.  These plots highlight the differences between the
$\alpha$-element abundances observed in different environments, which
are labelled in the upper plot (where gc stands for globular
clusters).
}
\label{alf1}
\end{figure}								     

\begin{figure}
\centerline{\hbox{\psfig{figure=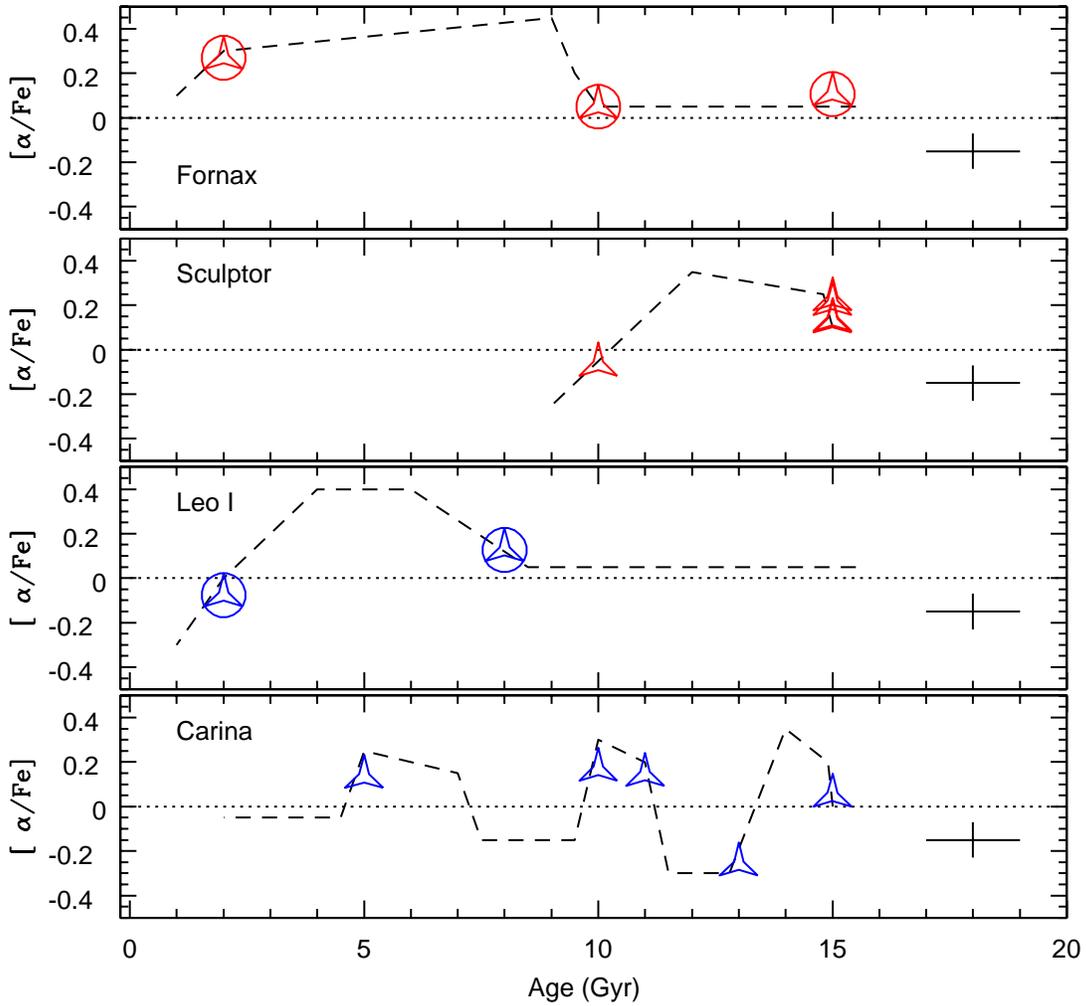,width=15cm}}}
\caption{ 
Here we plot an {\it illustrative} scenario which might allow us to
tie in our determinations of star formation history with
$\alpha$-element abundances for each galaxy.  The symbols are defined
as in Figure~15, and representative error bars are plotted. It is
obvious that the dashed lines cannot be constrained with the few data
we have.
}
\label{alf5}
\end{figure}								     

\begin{figure}
\centerline{\hbox{\psfig{figure=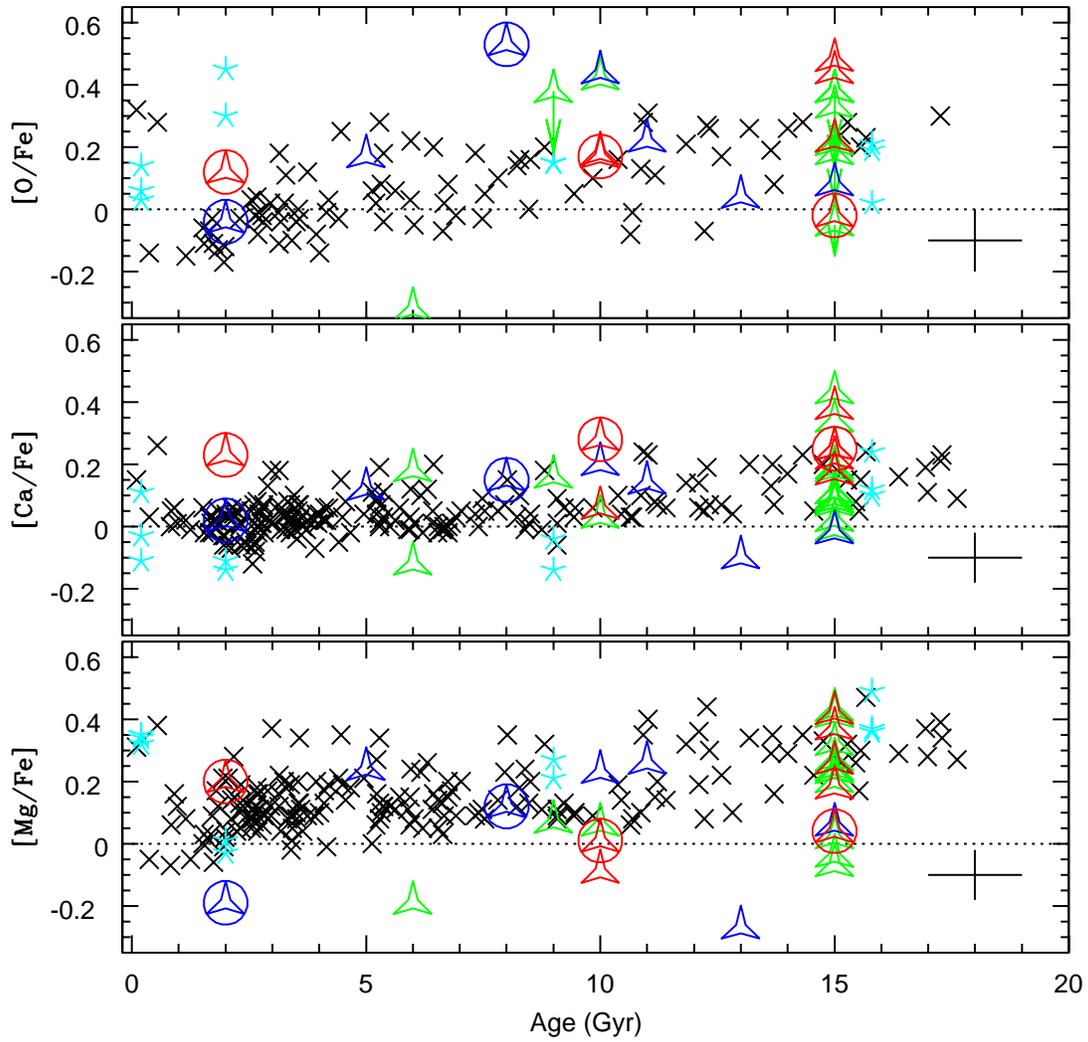,width=15cm}}}
\caption{
$\alpha$-Element abundances: Oxygen, Calcium \& Magnesium.  Our UVES
abundances are plotted versus age determined from isochrone analysis
described in $\S$3.2.  The symbols are defined as in Figure~15.
}
\label{alf2}
\end{figure}								     

\begin{figure}
\centerline{\hbox{\psfig{figure=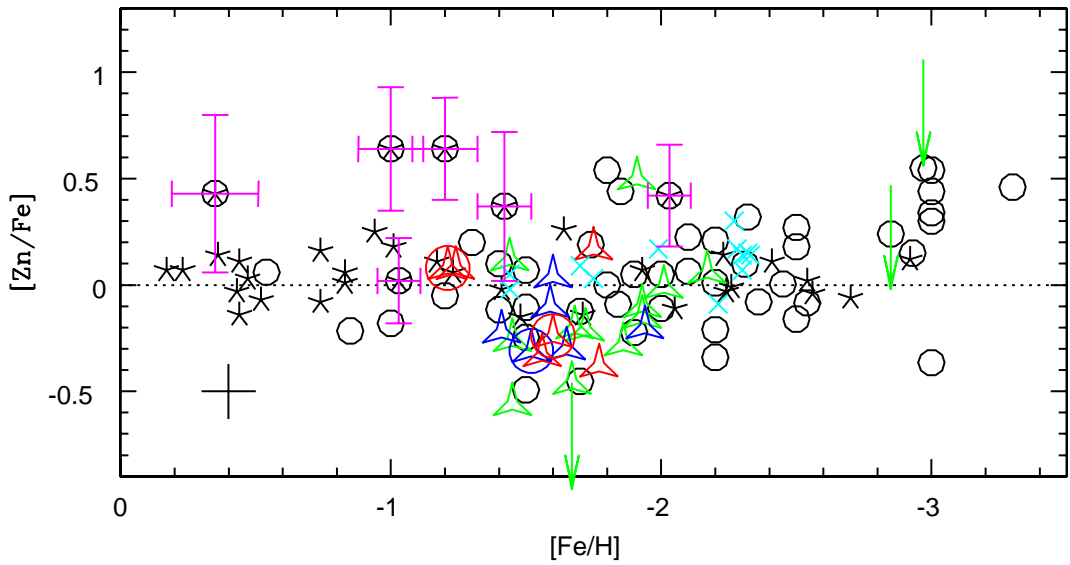,width=15cm}}}
\caption{Iron-Peak Abundances: Zinc.
[Zn/Fe] in the dSphs (symbols as in Figure 15) compared to abundances
in DLAs ({\it open circles} with error bars) from Pettini {\it et al.}
(2000), and Galactic halo stars ({\it stars} from Sneden {\it et al.}
1991; {\it small open circles} without error bars from Primas {\it et
al.} 2000).  A representative error bar for the stellar abundances is
shown in the lower left.
}
\label{zncu}
\end{figure}								     

\begin{figure}
\centerline{\hbox{\psfig{figure=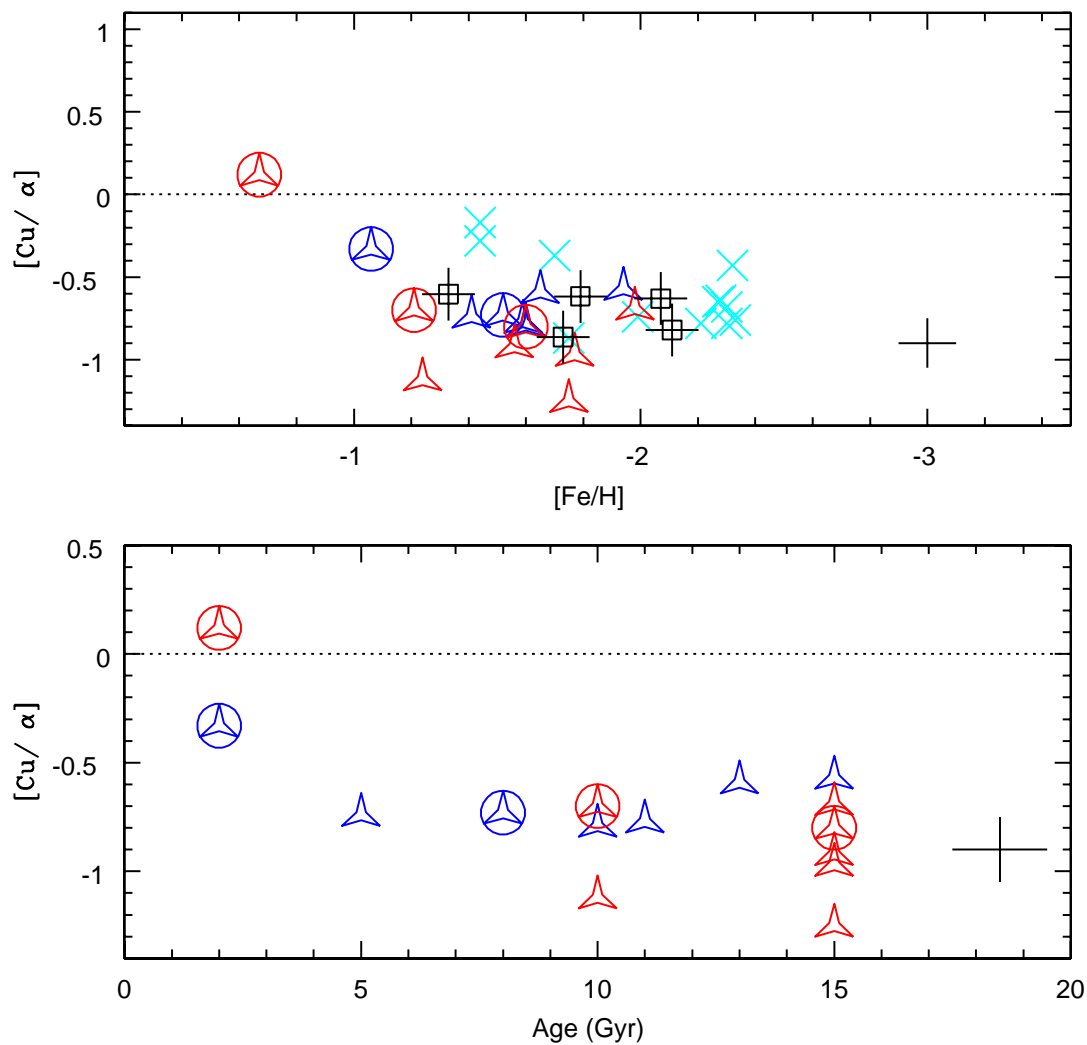,width=15cm}}}
\caption{
[Cu/alpha] versus [Fe/H] (top panel) and age (lower panel) in our dSph
stars to examine the nucleosynthetic source for Cu.  Since Cu/alpha is
quite flat over a wide range in metallicity and age, then we suggest
that Cu is not a SN Ia product (see Paper~I for further discussion).
The dSph symbols are the same as in Figure 15, with the addition of
{\it open squares} with error bars for Galactic halo stars from
Gratton \& Sneden (1988).  The large cross symbols are our globular
cluster measurements.  Representative error bars are shown in the
lower right of both panels.
}
\label{cualf}
\end{figure}								     

\begin{figure}
\centerline{\hbox{\psfig{figure=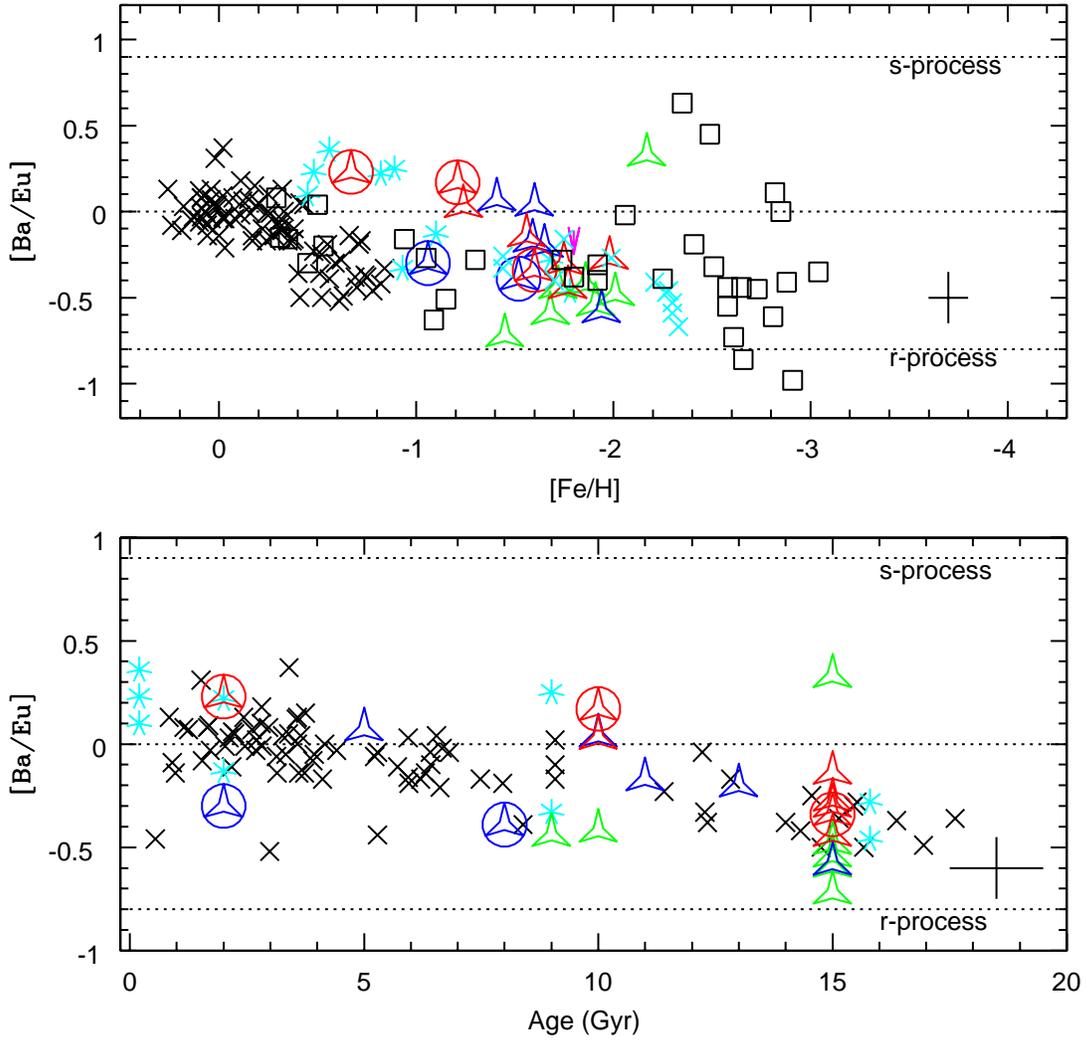,width=15cm}}}
\caption{Heavy Element Abundances: Barium \& Europium. 
[Ba/Eu] versus [Fe/H] (top panel) and age (lower panel) in our dSph
stars to examine contributions from the s-process by AGB stars,
particularly at intermediate and young ages.  Symbols are the same as
in Figure 15.  Representative error bars are shown in the lower right
of both panels.
}
\label{baeu}
\end{figure}

\begin{figure}
\centerline{\hbox{\psfig{figure=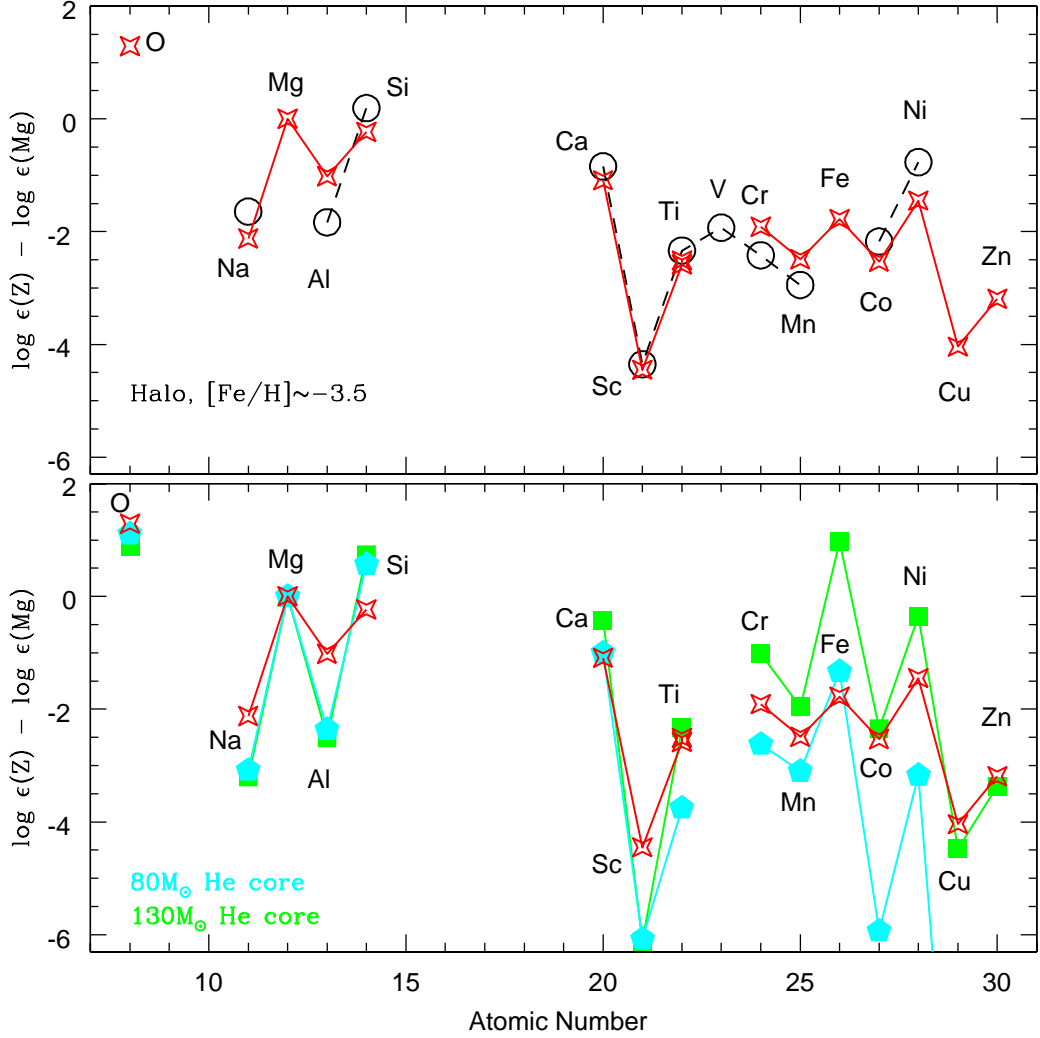,width=15cm}}}
\caption{
In the lower panel we show two comparisons with massive star
nucleosynthesis yield predictions of Heger \& Woosley (2002) and the
implied yields from our abundance measurements for the oldest stars in
the Sculptor dSph galaxy. In the upper panel we show in the same
manner the comparison between the old stars in Sculptor and an average
of the most metal poor stars observed in the Galactic halo ($-4 <$
[Fe/H] $< -3$) taken from McWilliam et al. (1995).  All the abundances
plotted here are normalized to Mg.
}
\label{cosmo}
\end{figure}								     
\end{document}